\documentstyle[preprint,eqsecnum,aps]{revtex}

\begin{document}

\draft

%\twocolumn[\hsize\textwidth\columnwidth\hsize\csname @twocolumnfalse\endcsname

\title{Linear Response of Thin Superconductors in Perpendicular
Magnetic Fields: An Asymptotic Analysis}

\author{Alan T. Dorsey\cite{atdemail}}

\address{Department of Physics,  University of Virginia, McCormick Road,
Charlottesville, VA 22901}

\date{\today}

\maketitle

\begin{abstract}

The linear response of a thin superconducting strip subjected
to an applied perpendicular time-dependent magnetic field
is treated analytically using the method of matched asymptotic
expansions.  The calculation of the induced current density is divided into
two parts: an ``outer'' problem, in the middle of the strip, which can
be solved using conformal mapping; and an ``inner'' problem near each of
the two edges, which can be solved using the Wiener-Hopf method.  The inner and
outer solutions are matched together to produce a solution which is
uniformly valid across the entire strip, in the limit that the
effective screening length $\lambda_{\rm eff}$  is small compared to the
strip width $2a$.
 From the current density it is shown that the perpendicular component
of the magnetic field inside the strip has a weak logarithmic singularity
at the edges of the strip.
The linear Ohmic response, which would be realized in a type-II superconductor
in the flux-flow regime, is calculated for both a  sudden jump in the
magnetic field and for an ac magnetic field. After a jump in the field
the current propagates in from the edges at a constant velocity $v=0.772 D/d$
(with $D$ the diffusion constant and $d$ the film thickness), rather than
diffusively, as it would for a thick sample.   The ac current density and the
high
frequency ac magnetization are also calculated.  The long time relaxation of
the current density after a jump in the field is found to decay exponentially
with
a time constant $\tau_{0} = 0.255 ad/D$.
The method is extended to treat the response of thin superconducting
disks, and thin strips with an applied current.  There is generally excellent
agreement between the results of the asymptotic analysis and the recent
numerical
calculations by E. H. Brandt [Phys. Rev. B {\bf 49}, 9024; {\bf 50}, 4034
(1994)].

\end{abstract}


% \vskip2pc]

\section{Introduction}

When a thin superconductor is placed in a perpendicular magnetic field there
are large demagnetizing fields which produce an enhanced response to the
applied field.  For instance, the induced magnetic moment in the perpendicular
geometry for a sample with thickness $d$ and a width $a\gg d$ is $O(a/d)$,
while
in for a longitudinal magnetic field it is only $O(1)$ \cite{brandt941}.
These demagnetizing fields are important not only for static properties,
but also for dynamic properties, such as the response of the sample to
an applied time-dependent magnetic field.  Surprisingly, there has been
little theoretical work on this subject until recently.   Brandt
has devised an efficient numerical method for calculating the
linear or nonlinear response of superconducting strips \cite{brandt941} or
disks \cite{brandt942}
in time-dependent magnetic fields.  His studies have unveiled a number
of interesting properties of the sheet current and magnetic moment in these
geometries, including dynamic scaling properties of the current density
at the sample edges at high frequencies or short times.

The present paper is an analytic treatment of the linear electrodynamics
of thin superconducting strips and disks in perpendicular magnetic fields, and
as such
is complementary to  Brandt's numerical work.   Many of the features
of the linear response which were extracted numerically by Brandt
emerge naturally from this analysis; the analytic results obtained here agree
quite well with Brandt's numerical results.  The calculational technique
employed here is the {\it method of matched asymptotic expansions}
\cite{vandyke,bender}, a
technique originally devised to treat boundary layer problems in
fluid mechanics \cite{vandyke}, and used recently to study
several interesting problems in nonequilibrium and inhomogeneous
superconductivity \cite{dorsey94,chapman94}.  The idea is to split
the problem into two pieces; an ``outer problem'' in the middle of the strip
(which is straightforward to solve), and an ``inner problem'' near each of the
two edges of the strip (which requires a bit more ingenuity).  The solutions
are then matched together,
and a ``uniform'' solution, valid across the entire strip, is constructed.
The time dependence of the applied field is treated using Laplace transforms,
which provides a unified framework for treating the transient response
after the applied field is suddenly switched on, as well as the
steady-state ac response.  It should be noted that there is an allusion
to such a matching procedure for the same problem in a paper by Larkin and
Ovchinnikov
\cite{larkin72}; these authors simply quote a result for the behavior of the
current near
the edge of a strip in a perpendicular field. The present work  goes well
beyond
that of Larkin and Ovchinnikov,  by explicitly constructing the inner, outer,
and uniform solutions, and using these solutions to study the {\it
nonequilibrium
response}.  It also appears that the result quoted by Larkin and Ovchinnikov
is incorrect in detail (see Appendix B).

As this paper is somewhat long, the primary results are collected in
Table \ref{table1}.  The organization of the paper is as follows.
In Sec. II the integro-differential equation for the current density in the
strip is derived, and a small parameter $\epsilon= 2 \lambda_{\rm eff}/a$
is identified, with $\lambda_{\rm eff}$ the effective penetration depth
and $2a$ the strip width.  Sec. III treats the large-$\epsilon$ limit,
which is helpful for anticipating some of the
features of the  small-$\epsilon$ solution.  The asymptotic analysis
for small-$\epsilon$ is constructed in Sec. IV; the outer problem is
essentially solved by conformal mapping, while the inner problem is solved
using
the Wiener-Hopf method.  The solutions are matched using formal
asymptotic matching, and the uniform solution is constructed.
The uniform solution is used to study the perpendicular magnetic field
in Sec. V, and the current density and magnetization in Sec. VI.
In Sec. VII the method is extended to treat the response of thin
superconducting disks, and thin strips in the presence of an applied
current. In Sec. VIII the long-time behavior is studied via an eigenfunction
expansion
of the current density, and the fundamental relaxation time is calculated
using a variation on the analysis of the previous sections.
The results are summarized in Sec. IX.  Some of the
more algebra intensive parts of the calculation are relegated to
Appendices A--C.

\section{Derivation of the Integro-differential Equation}

To begin, we will derive the integro-differential equation which
determines the current distribution in the strip.  The geometry is
illustrated in Fig. \ref{geometry}.  The strip is in the $x-y$ plane, being
infinite in the $y$-direction, and having a width $2a$ such that the
strip occupies the region $-a<x<a$.   The applied field
${\bf H}_{a}(t)  = H_{a}(t) \hat{z}$ is normal to the strip and in the
$z$-direction.  The vector potential ${\bf A}$ satisfies
\begin{equation}
-\nabla^{2} {\bf A} = 4\pi {\bf J},
\label{diffeq1}
\end{equation}
in the transverse gauge in which $\nabla\cdot {\bf A}=0$.
We will focus here on situations in which the current density  is invariant
along the $y$-direction (along the length of the strip), so that
both ${\bf J}$ and ${\bf A}$ are along the $y$-direction. In the thin
film approximation, the current density $J_{y} (x,z,t)$ is averaged over the
thickness $d$ of the film; the averaged current will be denoted
by $j(x,t)$, so we have
\begin{equation}
J_{y}(x,z,t) = d\, j(x,t)  \delta(z) \theta(a^2 - x^2).
\label{diffeq2}
\end{equation}
Eq. (\ref{diffeq1}) is solved by introducing the
Green's function for the two dimensional Laplacian, $G(x-x',z-z')$:
\begin{eqnarray}
A_{y}(x,z,t) & = &  A_{0,y} -4\pi  \int
     G(x-x',z-z') J_{y}(x,z,t)\, dx'\,dz' \nonumber \\
            & = &  A_{0,y} - 4\pi d \int_{-a}^{a} G(x-x',z) j(x',t)\, dx'.
\label{diffeq3}
\end{eqnarray}
Differentiating both sides with respect to $x$,  and using
$\partial A_{0,y}/\partial x = H_{a}(t)$, we obtain
\begin{eqnarray}
  H_{z} (x,z,t) & = & {\partial A_{y}(x,z,t) \over \partial x} \nonumber \\
         & = &  H_{a}(t) -4\pi d \int_{-a}^{a}
       {\partial G(x-x',z) \over \partial x} j(x',t)\, dx'.
\label{diffeq4}
\end{eqnarray}
Finally, if we specialize Eq. (\ref{diffeq4}) to $z=0$,  and use
$\partial G(x-x',0)/\partial x = 1/2\pi(x-x')$, we obtain for the magnetic
field normal to the strip,
\begin{equation}
H_{z} (x,t) = H_{a}(t)   + 2 d (P) \int_{-a}^{a}
       {j(x',t) \over x' - x} \, dx',
\label{diffeq5}
\end{equation}
where $(P)$ indicates a principle value integral.

To complete the description, we require a constitutive relation between
the averaged current and the fields.   This paper will concentrate on the
linear response of the current,  so that the general time-dependent
response is
\begin{equation}
 j(x,t) = \int_{-\infty}^{t} \sigma(t-t') E_{y}(x,t') \, dt',
\label{ohm}
\end{equation}
with $\sigma$ the conductivity.     In most of this paper we shall
be interested in the solution of the initial value problem; i.e., the
time evolution of the sheet current after an applied current has been
switched on.  Then it is natural to Laplace transform the currents and
the fields with respect to time $t$:
\begin{equation}
j(x,s)  = \int_{0}^{\infty}  e^{-st} j(x,t)\, dt,
\label{laplace}
\end{equation}
and so on.  The inverse Laplace transform is
\begin{equation}
j(x,t) = {1\over 2\pi i} \int_{c-i\infty}^{c+i\infty}
                     e^{st} j(x,s) \,ds,
\label{invert}
\end{equation}
with the integration contour chosen to pass to the right of any
singularities.
After Laplace transforming, Eq.\ (\ref{current1}) becomes
\begin{eqnarray}
 j(x,s) &=&  \sigma(s) E_{y}(x,s) \nonumber \\
        & = & - s \sigma(s) A_{y}(x,s),
\label{ohm2}
\end{eqnarray}
where $E_{y}(x,t) = - \partial A_{y}(x,t)/\partial t$ and an integration by
parts has been used to
obtain the last line.  Now Laplace transform Eq.\ (\ref{diffeq5})
with respect to time, and use Eq.\ (\ref{ohm2}) to eliminate the
fields in favor of the currents, to obtain
the following equation of motion for the current:
\begin{equation}
- 4\pi\lambda_{\rm eff}(s) d\, {\partial j(x,s) \over \partial x}
 = H_{a}(s) + 2 d \,(P) \int_{-a}^{a} { j(x',s)\over x'-x}\, dx',
\label{current1}
\end{equation}
where $\lambda_{\rm eff}(s)$ is an effective screening length defined
by
\begin{equation}
\lambda_{\rm eff} (s) = {1\over 4\pi d s\sigma(s)}.
\label{lambda}
\end{equation}
For a superconductor, $\sigma(s) = 1/4\pi \lambda^{2} s$, with $\lambda$ the
London penetration depth, so that in this case
$\lambda_{\rm eff} = \lambda^{2}/d$.  For an Ohmic conductor,
$\lambda_{\rm eff} = 1/4\pi \sigma(0)d s = D/ds$, with $D$ the diffusion
constant for the magnetic flux.  An Ohmic response would be realized in a
normal metal
or in a type-II superconductor in the flux-flow regime. Eq. (\ref{current1}) is
essentially identical to the equations of motion derived by Larkin and
Ovchinnikov
\cite{larkin72}, Eq. (33),  and Brandt \cite{brandt941}, Eq. (3.6)
the only difference being that Laplace rather than Fourier transforms
are being used here.
This equation of motion can be conveniently expressed in dimensionless
variables by writing  $x'=x/a$, $f(x',s) = j(x,s)/(H_{a}(s)/2\pi d)$, and
$\epsilon= 2\lambda_{\rm eff}(s)/a$:
\begin{equation}
- \epsilon f'(x,s) = 1  + {1\over \pi} (P) \int_{-1}^{1}
            {f(x',s)\over x'-x}\, dx'
\label{current2}
\end{equation}
(the $s$ dependence of $\epsilon$ will be suppressed for notational
simplicity; it will be reinstated later when the transforms are
inverted).  The current density is the primary quantity of interest
in this paper.  A related, and experimentally accessible, quantity
is the magnetic moment per unit length of the strip \cite{brandt941},
\begin{eqnarray}
M(s) & = & \int_{-a}^{a} x\, j(x,s) \, dx \nonumber \\
   & = & -  {a^2 H_{a} \over 4 d } m(s)
\label{momentdefine}
\end{eqnarray}
where $m(s)$ is a dimensionless moment defined as
\begin{equation}
 m(s) =  - {2 h(s) \over \pi} \int_{-1}^{1} x f(x,s)\, dx,
\label{ms}
\end{equation}
where $h(s)$ is defined through $H_{a}(s) = H_{a} h(s)$.

There are no known analytical solutions of Eq.\ (\ref{current2}); its
solution is the subject of the remainder of this paper.   However, for
a typical sample $\epsilon \ll 1$, in which case the left hand side
of Eq. (\ref{current2}) constitutes a singular perturbation (the highest
derivative is multiplied by the small parameter), and we
can bring to bear all of the techniques of asymptotic analysis to
solve this problem \cite{bender}.   Before attempting this, we will
first develop
a perturbative analysis which is valid for $\epsilon \gg 1$, which
is physically less interesting but mathematically simpler.

\section{Expansion for Large $\epsilon$}

To study the behavior of Eq. (\ref{current2}) for large $\epsilon$, it is
useful to first rescale by introducing a new function
$g(x,s) = \epsilon f(x,s)$.
We then expand $g(x,s)$ in powers of $\epsilon^{-1}$:
\begin{equation}
g(x,s) \sim g_{0}(x,s) + \epsilon^{-1} g_{1}(x,s) + \ldots.
\label{large_eps}
\end{equation}
Substituting into Eq. (\ref{current2}) and matching terms of the same order,
we have
\begin{equation}
g_{0}'(x,s) = -1,
\label{goeqn}
\end{equation}
\begin{equation}
 g_{1}'(x,s) = - {1\over \pi}  (P) \int_{-1}^{1} { g_{0}(x',s) \over x'-x}\,
dx',
\label{g1eqn}
\end{equation}
and so on with the higher order terms.  Assuming that there is no net
current in the strip (i.e., no applied current), then the  current is odd
in $x$, so that $g(0) = 0$, and we have
\begin{equation}
g_{0}(x,s) = - x,
\label{g0}
\end{equation}
\begin{equation}
g_{1}(x,s) = {1\over \pi} \left[ x + { 1-x^{2} \over 2}
           \ln \left( { 1+x \over 1-x}\right) \right].
\label{g1}
\end{equation}
The current increases linearly across the sample,
except near the edges.  Near the left edge ($x=-1$), we have
\begin{equation}
g(x,s)\sim  1 - {1\over\pi\epsilon} \left[ 1 + (1+x)\ln(2/(1+x)) \right].
\label{g2}
\end{equation}
As we shall see below, a similar behavior near the edge
will also emerge in the small-$\epsilon$ limit.    From the
expansion for $f(x,s)$ we can calculate the
magnetic moment from Eq. (\ref{ms}):
\begin{equation}
m(s) =  h(s)\left[ {4\over 3} {1\over \pi \epsilon(s) }
       - {2\over (\pi \epsilon(s))^{2}} + O(\epsilon^{-3}) \right] .
\label{largeepsilon}
\end{equation}
For an Ohmic conductor, $\epsilon(s) = 2D/ads$, so the magnetic
moment vanishes linearly with frequency at low frequency.
This result is essentially equivalent to Eq. (5.6) of
Ref. \cite{brandt941}.

\section{Asymptotic Analysis for Small $\epsilon$}

We now turn to the solution of Eq. (\ref{current2}) for small $\epsilon$.
An asymptotic solution can be obtained using the method of matched
asymptotic expansions.  We break up the strip into an ``outer region,''
which is the interior of the strip, and two ``inner regions,'' one near
each edge.  This is illustrated schematically in Fig. \ref{schematic}. The
solutions are then matched in common overlap regions.
A detailed discussion of the method can be found in Ref. \cite{bender}.

\subsection{Outer solution}

The expansion in the outer region (away from the edges) is obtained
by expanding $f$ as a series in $\epsilon$:
\begin{equation}
f(x;\epsilon) \sim f_{0}(x) + \epsilon f_{1}(x) + \ldots \ .
\label{asmpt1}
\end{equation}
Substituting this expansion into Eq. (\ref{current2}) and collecting
terms of the same order, we find for $f_{0}(x)$
\begin{equation}
  {1\over \pi}  (P) \int_{-1}^{1} {f_{0}(x')\over x'-x}\, dx' = -1.
\label{cauchy}
\end{equation}
The solution of this singular integral equation which is odd is $x$
can be found in
Ref. \cite{CKP}:
\begin{equation}
 f_{0}(x) =  - {x\over (1-x^2)^{1/2}},
\label{outer2}
\end{equation}
which coincides with the usual solution obtained from conformal mapping
techniques.  Near the edges at $x=\pm 1$, $f_{0}$ behaves as
\begin{equation}
f_{0}\sim \mp {1 \over [2(1\mp x)]^{1/2}},
\label{outerlim}
\end{equation}
so the current has a square root divergence at the edges.  This is due to
the fact that the outer solution corresponds to complete screening
of the applied field, which can only be achieved by having an infinite
current density at the edges.  The outer solution therefore breaks down
at distances of order $\epsilon$ of the edges; the current at the
edges is thus of order $\epsilon^{-1/2}$.   In order to remedy this
problem, we proceed to the solution of the inner problem near each of
the edges.

\subsection{Inner solution}

We will first study the inner problem at the left edge, $x=-1$.  The
outer solution breaks down at $x+1\sim \epsilon$, suggesting that the
appropriate variable in the inner region is
\begin{equation}
X= (x+1)/\epsilon.
\label{inner11}
\end{equation}
Also, since $f(-1)\sim \epsilon^{-1/2}$, we rescale $f(x)$ in the inner
region as
\begin{equation}
 F(X) = \epsilon^{1/2} f(x),
\label{inner12}
\end{equation}
so that $F(0)=O(1)$.
In terms of these inner variables, Eq. (\ref{current2}) becomes
\begin{equation}
F' = -\epsilon^{1/2} - {1\over \pi} (P) \int_{0}^{2/\epsilon}
    {F(X')\over X'-X}\, dX'.
\label{inner2}
\end{equation}
Next, expand $F(X;\epsilon)$ in powers of $\epsilon^{1/2}$, as
suggested by the rescaled form of the integro-differential equation:
\begin{equation}
F(X;\epsilon) \sim F_{0}(X) + \epsilon^{1/2} F_{1}(X) + \ldots \ .
\label{inner3}
\end{equation}
The lowest order term satisfies
\begin{eqnarray}
F_{0}'(X) & = & - {1\over \pi} (P) \int_{0}^{2/\epsilon}
{F_{0}(X')\over (X'-X)} dX' \nonumber \\
  & = & - {1\over \pi} (P) \int_{0}^{\infty}
{F_{0}(X')\over (X'-X)} dX'  + {1\over \pi} (P) \int_{2/\epsilon}^{\infty}
{F_{0}(X')\over (X'-X)} dX'.
\label{inner3a}
\end{eqnarray}
For small $\epsilon$ the second integral will be dominated by the
large $X$ behavior of $F_{0}(X)$; it will be shown below that in order
to match onto the outer solution this is necessarily of the form
$F_{0}(X)\sim (2X)^{-1/2}$.  Therefore, we see that the second integral
is of order $\epsilon^{1/2}$, and can be dropped at this order
of the calculation.  Our final integral equation for $F_{0}(X)$ is
then
\begin{eqnarray}
F_{0}'(X) = - {1\over \pi} (P) \int_{0}^{\infty}
{F_{0}(X')\over (X'-X)} dX'.
\label{inner4}
\end{eqnarray}
The problem in the inner region consists of solving a homogeneous
integro-differential equation on a semi-infinite interval.
This is equivalent to finding the current
distribution in a semi-infinite strip in zero applied magnetic field.

Eq. (\ref{inner4}) can be solved using the Wiener-Hopf method \cite{CKP},
as follows.  The function $F_{0}(X)\rightarrow 0$ as $X\rightarrow \infty$, and
$F_{0}(X)=0$ for $X<0$; introduce a second unknown function $G(X)$ such that
$G(X)=0$ for $X>0$ and $G(X)\rightarrow 0 $ as $X\rightarrow -\infty$.
We then introduce the complex Fourier transforms of $F_{0}(X)$ and $G(X)$,
\begin{equation}
\Phi_{+}(k) = \int_{0}^{\infty} F_{0}(X) e^{ikX} dX,
\label{fourier1}
\end{equation}
\begin{equation}
G_{-}(k) = \int_{-\infty}^{0} G(X) e^{ikX} dX,
\label{fourier2}
\end{equation}
such that $\Phi_{+}(k)$ is analytic for ${\rm Im}(k)>-\beta$ and
$G_{-}(k)$ is analytic for ${\rm Im}(k)<\alpha$, for some $\alpha>\beta$.
Then Fourier transforming Eq.\ (\ref{inner4}), and integrating by parts,
we obtain
\begin{equation}
 G_{-}(k)  - F_{0}(0) - i k \Phi_{+}(k) =  i {\rm sgn}(k)  \Phi_{+}(k).
\label{WH1}
\end{equation}
We see that as $k\rightarrow \infty$,
\begin{equation}
\Phi_{+}(k) \sim - {F_{0}(0)\over ik} + O(k^{-2}).
\label{WH2}
\end{equation}
To take care of the ambiguities in defining ${\rm sgn}(k)$, replace it by
$k/(k^{2} + \delta^{2})^{1/2}$, with the real part $>0$ for ${\rm Re}(k)>0$,
and choose the branch cuts to run between
$(-i\infty, -i\delta)$ and $(i\delta,i\infty)$.   We can then take
$\delta\rightarrow 0$ at some later point in the calculation.  Rearranging
Eq.\ (\ref{WH1}) a bit, we have
\begin{equation}
 ik  K(k) \Phi_{+}(k)
   =   G_{-}(k) - F_{0}(0),
\label{WH3}
\end{equation}
where
\begin{equation}
 K(k) = 1 + {1\over (k^{2} + \delta^{2})^{1/2}}.
\label{kernel}
\end{equation}
Now, if we can factor $K(k)$ into the form  $K(k) = K_{+}(k)/K_{-}(k)$,
with $K_{+}(k)$ analytic for ${\rm Im}(k)>-\delta$ and $K_{-}(k)$ analytic
for ${\rm Im}(k)<\delta$, then we may rewrite Eq.\ (\ref{WH3}) as
\begin{equation}
 ik  K_{+} (k) \Phi_{+}(k)
   = K_{-}(k)  \left[ G_{-}(k) - F_{0}(0) \right].
\label{WH5}
\end{equation}
Both sides are now analytic in their respective regions of analyticity; we
can then use analytic continuation arguments to note that both sides must then
equal an entire function $E(k)$.   By examining the limiting behavior
of the left hand side as $k\rightarrow \infty$, we see that this
function must be chosen to be a constant $C$ (any positive power of $k$
would produce non-integrable singularities in $F_{0}(X)$),
so that we finally have
\begin{equation}
\Phi_{+}(k) =  { C \over ik  K_{+}(k) }.
\label{WH6}
\end{equation}
The function $F_{0}(X)$ is then obtained by inverting the Fourier
transform,
\begin{equation}
 F_{0} (X) = {1\over 2 \pi} \int_{-\infty}^{\infty}  \Phi_{+}(k) e^{-ikX} dk,
\label{inverse}
\end{equation}
with the integration path indented so as to pass above any singularities
on the real axis.

The only remaining task is the decomposition of $K(k)$, which is carried out
in Appendix A.   Using Eqs. (\ref{simple2}) and (\ref{simple3}), we have
\begin{eqnarray}
F_{0}(X) & = &  {C\over 2\pi  i} \int_{-\infty}^{\infty}
   {e^{-i\varphi(k)} \over k (1 + 1/|k|)^{1/2}} e^{-ikX}\, dk \nonumber \\
   & = & - {C\over \pi} \int_{0}^{\infty}
   { \sin\left[ kX + \varphi(k)\right] \over (k^2 + k)^{1/2} } \, dk,
\label{F01}
\end{eqnarray}
where the phase $\varphi(k)$ is given by
\begin{equation}
\varphi (k) = {\pi \over 4} + {1\over \pi} \int_{0}^{k} {\ln u \over 1-u^2}
            \, du.
\label{F02}
\end{equation}
The constant $C$ is determined from the matching conditions, which are
discussed below.

\subsection{Asymptotic matching}

The inner and outer solutions may now be matched together in a suitable
overlap region.  This is done by expressing the outer solution $f_{0}(x)$
in terms of the inner variable $X=(x+1)/\epsilon$, and then taking
$X\rightarrow 0$ while holding $\epsilon$ fixed:
\begin{eqnarray}
f_{0}(X) &=& -{\epsilon X - 1 \over
                 [ (2-\epsilon X) \epsilon X]^{1/2}}\nonumber \\
         &\sim &   {1 \over (2\epsilon X)^{1/2}} \qquad (X\rightarrow 0).
\label{match1}
\end{eqnarray}
The inner solution $F_{0}(X)$ must match onto this outer solution as
$X\rightarrow \infty$, so the asymptotic behavior of $F_{0}(X)$
must be
\begin{equation}
F_{0}(X) \sim  {1 \over (2X)^{1/2}} \qquad (X\rightarrow \infty).
\label{match2}
\end{equation}
Now expand Eq.\ (\ref{F01}) for large-$X$; the integral is dominated
by the small-$k$ behavior of the integrand, and we find
\begin{equation}
F_{0}(X) \sim -  {C\over (\pi X)^{1/2}},
\label{largeX}
\end{equation}
so that we have
\begin{equation}
C = - \left( {\pi \over 2}\right)^{1/2}
\label{match3}
\end{equation}
in order to match the inner and outer solutions.  Therefore our final
expression for the inner solution is
\begin{equation}
F_{0}(X) = {1 \over (2\pi)^{1/2} } \int_{0}^{\infty}
   { \sin\left[ kX + \varphi(k)\right] \over (k^2 + k)^{1/2} } \, dk.
\label{final}
\end{equation}
Comparing Eqs.\ (\ref{WH2}) and (\ref{WH6}), we see that
$F_{0}(0) = - C = (\pi/2)^{1/2}$.

By rotating the integration contour, it is possible to show that for
$X<0$ the integral vanishes (as it should), while for $X>0$ the integral
may be written as
\begin{equation}
F_{0}(X) = {1\over (2\pi)^{1/2}} \int_{0}^{\infty} {e^{-Xy - g(y)}
 \over y^{1/2} (y^{2} +1)^{3/4} } \, dy,
\label{rotate1}
\end{equation}
with
\begin{equation}
 g(y) = {1\over \pi} \int_{0}^{y} {\ln u \over 1 + u^2 }\, du.
\label{rotate2}
\end{equation}
This function has the limiting behaviors
\begin{eqnarray}
 g(y) = \left\{
     \begin{array}{ll}
     -  y \ln(e/y)/\pi  + O(y^3), \qquad y \ll 1 ; \\
    - \ln(ey)/\pi y + O(y^{-3}), \qquad y\gg 1.
     \end{array}
  \right.
\label{g(y)}
\end{eqnarray}
The square-root singularity in the integrand at $y=0$ can be
removed by changing variables to $y=z^2$, so that
\begin{equation}
F_{0}(X) = \left( {2\over \pi}\right)^{1/2} \int_{0}^{\infty}
 { e^{-X z^2 - g(z^2)} \over (z^4 + 1)^{3/4}}\, dz,
\label{rotate3}
\end{equation}
which is particularly convenient for numerical evaluation.
As shown in Appendix B, for small-$X$, $F_{0}(X)$ has the expansion
\begin{equation}
F_{0}(X) = {1\over (2\pi)^{1/2}}\left[\pi - 1.4228 X + X\ln X + O(X^2)\right]
\label{smallX}
\end{equation}
(which disagrees with Eq.\ (37) of Larkin and Ovchinnikov,
Ref.\ \cite{larkin72}; see the remarks in Appendix B).  Although the
current is finite at $X=0$, it has a slope which diverges as $\ln(X)$.
The result of a numerical
evaluation of the integral, along with a comparison of the numerical results
to the asymptotic expansions,  is shown in Fig.\ \ref{inner_fig}.

So far we have only discussed the matching procedure at the left edge
of the strip $x=-1$.  The same procedure can be carried out at the
right edge, $x=1$, as follows.  At the right edge, the inner
variable will be $\bar{X}=(1-x)/\epsilon$. As before, we will also
rescale $f(x)$ as $\bar{F}(\bar{X}) = \epsilon^{1/2} f(x)$.
Substituting these expressions into Eq.\ (\ref{current2}), we obtain
\begin{equation}
\bar{F}'(\bar{X}) = \epsilon^{1/2} - {1\over \pi} (P) \int_{0}^{2/\epsilon}
    {\bar{F}(\bar{X}')\over \bar{X}'-\bar{X}}\, d\bar{X}',
\label{right1}
\end{equation}
which is the same as Eq.\ (\ref{inner2}) except for the minus sign in
front of the $\epsilon^{1/2}$.  Expanding in powers of $\epsilon^{1/2}$,
the $O(1)$ term, $\bar{F}_{0}(\bar{X})$, satisfies Eq.\ (\ref{current2}),
and the method of solution is identical.  To match onto the outer solution,
we write $f_{0}(x)$ in terms of $\bar{X}$, and then take
$\bar{X}\rightarrow 0$ while holding $\epsilon$ fixed:
\begin{eqnarray}
f_{0}(\bar{X}) &=& - {1- \epsilon \bar{X}  \over [ (2-\epsilon \bar{X})
        \epsilon \bar{X}]^{1/2}}\nonumber
\\
         &\sim & -  {1 \over (2\epsilon \bar{X})^{1/2}}
                            \qquad (\bar{X}\rightarrow 0).
\label{right2}
\end{eqnarray}
Therefore, the asymptotic behavior of $\bar{F}_{0}(\bar{X})$ must be
\begin{equation}
\bar{F}_{0}(\bar{X}) \sim  -{1 \over (2\bar{X})^{1/2}} \qquad
          (\bar{X}\rightarrow \infty).
\label{right3}
\end{equation}
Therefore, $\bar{F}_{0}(\bar{X}) = - F_{0}(\bar{X})$, which could have
been surmised from the symmetry of the problem.

\subsection{Uniform solution}

We are now in a position to construct an asymptotic  solution which is
uniformly valid across the entire width of the strip; i.e., valid for all
$x$ as $\epsilon\rightarrow 0$.  To do this we simply add the inner and
outer solutions;  however, this would
produce a result which was $2f_{\rm match}(x)$ in the matching region,
so we also need to subtract the $f_{\rm match}(x)$ for each of the
two matching regions \cite{bender}.  The result is
\begin{eqnarray}
f_{\rm unif} (x,s)& = & -{x\over (1-x^2)^{1/2}}   + {1\over [2(1-x)]^{1/2}}
           - {1 \over [2(1+x)]^{1/2}}  \nonumber \\
       & & \qquad + \epsilon(s)^{-1/2} \left\{ F_{0}[(1+x)/\epsilon(s)]
           -  F_{0}[(1-x)/\epsilon(s)] \right\}.
\label{uniform}
\end{eqnarray}
The uniform solution is plotted in Fig. \ref{uniform_fig} for $\epsilon=0.1$.

With the uniform solution we can calculate the magnetic moment, given
in Eq.\ (\ref{ms}).  The result is
\begin{equation}
m(s) = h(s) \left\{ 1 - {8\over 3\pi} + \left( {2\over \pi} \right)^{3/2}
\int_{0}^{\infty} { e^{-g(\epsilon y)} [ (y+1)e^{-2y} + y -1] \over
           y^{5/2} [1 + \epsilon^{2} y^2]^{3/4} } \, dy \right\}.
\label{moment1}
\end{equation}
For $\epsilon = 0$, the integral is $8/3\pi$, so that $m(s) = h(s)$, which
is the ideal screening limit.  Determining the leading $\epsilon$ behavior
of the integral is rather subtle; the details are relegated to Appendix C.
The result is
\begin{equation}
m(s) = h(s) \left[ 1 - {6\over \pi^2} \ln\left( {8 e^{\gamma -5/3} \over
               \epsilon}\right) \epsilon \right].
\label{moment2}
\end{equation}
This expression should give the correct high-frequency ($s\rightarrow\infty$)
behavior of the magnetization.  Note that for $s=0$
($\epsilon\rightarrow \infty$),
$m(s)= 1- 8/3\pi = 0.151$; however, we know that the correct limiting
behavior is $m(0) = 0$ (see Sec. III).  Therefore
the uniform approximation does not reproduce the correct low frequency
behavior of the magnetization for an Ohmic conductor.

\section{Magnetic field within the strip}

By using the constitutive relation, Eq. (\ref{ohm2}), it is also
possible to calculate the magnetic field perpendicular to the
strip.  Using $H_{z}(x,s) = \partial A_{y}(x,s)/\partial x$,
going to our dimensionless variables, and using the uniform
approximation from the section above, we have
\begin{eqnarray}
H_{z}(x,s) & = & - H_{a}(s) \epsilon(s) {\partial f(x,s) \over \partial x}
                          \nonumber \\
 & = & H_{a}(s) \left\{ \epsilon(s) \left[ {1\over (1-x^2)^{3/2}}
  - {1\over [2(1-x)]^{3/2}} - {1\over [2(1+x)]^{3/2}} \right] \right.
    \nonumber \\
 & & \qquad \left.  - \epsilon(s)^{-1/2}\left\{ F_{0}'[ (1+x)/\epsilon(s)]
      + F_{0}'[ (1-x)/\epsilon(s)] \right\} \right\}.
\label{field}
\end{eqnarray}
The magnetic field is plotted in Fig. \ref{field_fig}.  From the results
in Appendix B, close to the edges ($1\pm x = O(\epsilon)$) we have
\begin{equation}
F_{0}'(X) = - {1\over (2\pi)^{1/2}} \ln(1/X) + O(1),
\label{field2}
\end{equation}
so that it would appear that the field diverges logarithmically
at the edges, in agreement with the numerical work in Ref. \cite{brandt941}.
However, as we get even closer to the edge ($1\pm x = O(\epsilon^{3})$),
this log divergence is swamped  by a square-root divergence from the
outer and overlap terms.  This latter behavior is most likely an
artifact of the approximation, and would probably disappear in a
higher-order calculation.

\section{Dynamics of the current density and magnetization}

Having obtained a uniformly asymptotic solution to the equation of motion
for the averaged current density, we can now examine its evolution in the
time domain by inverting the Laplace transform for $j(x,s)$,
Eq.\ (\ref{invert}).
The details of the inversion process will depend
upon the time dependence of the applied field, and the model chosen for
$\epsilon(s)$.  Two different models for
$\epsilon(s)$ will be considered: (1) $\epsilon(s) = 2\lambda^{2}/a d$
a constant, corresponding to a superconductor;
(2) $\epsilon(s)= (2 D/a d) (1/s)$, with $D= 1/4\pi\sigma(0)$ the
diffusion constant for flux in the normal phase, corresponding to
an Ohmic conductor (a type-II superconductor in the flux-flow
regime, for instance \cite{brandt941}).

\subsection{Superconductor}

For a superconductor the inversion of the Laplace transform is
particularly simple.  In conventional units we have (recall
that $\lambda_{\rm eff} = \lambda^2/d$)
\begin{eqnarray}
j_{\rm unif}(x,t) & = & {H_{a}(t)\over 2\pi d}
    \left\{ - {x\over (1 - x^2)^{1/2}}
  + \left[ {1 \over 2(1-x) } \right]^{1/2}
  - \left[ {1 \over 2(1+x) } \right]^{1/2} \right.\nonumber \\
  & & \qquad \left. \epsilon^{-1/2}
  \left[ F_{0}\left( {1+x\over \epsilon}\right)
    -  F_{0}\left( {1-x\over \epsilon} \right) \right] \right\}.
\label{superconductor}
\end{eqnarray}
In this case the induced current is in phase with the applied field.

\subsection{Ohmic conductor: penetration of a jump in the applied field}

 We will first treat the case in which the field
is suddenly switched on, so that $H_{a}(t) = H_{a} \theta(t)$,
and thus $H_{a}(s) = H_{a} /s$.  Inverting the Laplace transform, we
have
\begin{eqnarray}
j_{\rm unif}(x,t) & = & {H_{a}(t)\over 2\pi d} \left\{ - {x\over (1 -
x^2)^{1/2}}
  + \left[ {1 \over 2(1-x) } \right]^{1/2}
  - \left[ {1 \over 2(1+x) } \right]^{1/2} \right\} \nonumber \\
   & & \qquad + j_{\rm edge}(1+x,t) - j_{\rm edge}(1-x,t),
\label{jump1}
\end{eqnarray}
where the edge current $j_{\rm edge}(\bar{x},t)$ is given by
\begin{eqnarray}
j_{\rm edge} (\bar{x},t) & = & { H_{a} \over 2 \pi d}
  \left( {a d \over 4 D}\right)^{1/2} \int_{0}^{\infty}{ e^{- g(y)}
        \over y^{1/2} (y^{2} + 1)^{3/4} } \, dy \nonumber \\
   &  & \qquad \times {1\over 2\pi i}  \int_{c-i\infty}^{c+i\infty}
        {\exp[ s t - (a d \bar{x}/2 D) y s] \over (\pi s)^{1/2}}\, ds.
\label{edge1}
\end{eqnarray}
The Laplace transform can be calculated by closing the contour in the left
half plane (for $t>d \bar{x}/2 D$), wrapping the contour around the
branch cut along the negative $s$ axis, with the result
\begin{equation}
j_{\rm edge} (\bar{x},t) =  { H_{a} \over 2 \pi d}
     \left( {a d \over 4 D t}\right)^{1/2} {\cal F}_{1}\left( {ad\over 2D}
     {\bar{x}\over t} \right),
\label{edge2}
\end{equation}
where ${\cal F}_{1}(u)$ is a scaling function given by
\begin{equation}
{\cal F}_{1}(u) = {1\over \pi} \int_{0}^{1/u} { e^{-g(y)} \over (1-uy)^{1/2}
              y^{1/2} (y^2 +1)^{3/4} } \, dy.
\label{edge3}
\end{equation}
The scaling function has been normalized so that ${\cal F}_{1}(0) = 1$;
for large-$u$, ${\cal F}_{1}(u)\sim u^{-1/2}$.  For numerical purposes,
it is useful to transform the integral by making the substitution
$1-u y = \sin^{2}\theta$, so that
\begin{equation}
{\cal F}_{1}(u) = {2 u \over \pi} \int_{0}^{\pi/2}
{\exp[-g(\cos^{2}\theta /u)] \over (\cos^{4}\theta + u^{2})^{3/4}}\,
  d\theta,
\label{newedge}
\end{equation}
which removes the square-root singularities from the integrand at the
endpoints of integration.

The scaling function ${\cal F}_{1}(u)$ is plotted in Fig. \ref{scaling_fig}.
There is a maximum at $u=0.386$, with
${\cal F}_{\rm 1,max} = 1.1078$.  From the position of this peak we can
define a velocity $v$ of flux penetration:
\begin{equation}
v = \left( {\bar{x} \over t}\right)_{\rm max} =  0.772 {D\over d}.
\label{velocity}
\end{equation}
These results agree exactly with the numerical work of
Brandt \cite{brandt941,brandt942}.  As noted by Brandt, in the thin film
geometry the flux entry is ballistic rather than diffusive, as it
would be in a bulk sample with no demagnetizing fields.

We can also calculate the time dependent magnetization after a jump in the
field.  By using the small-$\epsilon$ expansion in Eq. (\ref{moment2}),
and inverting the Laplace transform by wrapping the integration contour
around the branch cut along the negative-$s$ axis, we find  at short times
\begin{equation}
m(t) = 1 - {6\over \pi^{3}} (t/\tau) \ln( 8\pi e^{-2/3} t/\tau) +  O(t^2),
\label{mt}
\end{equation}
where $\tau = ad/2\pi D$ is a characteristic relaxation time.
Similar behavior was found by Brandt \cite{brandt941} in his
numerical studies of the magnetization; for the prefactor of the
log he obtained $0.205$, compared to the present value of
$6/\pi^3 = 0.194$; for the constant inside the log, he obtained
$25$, compared to our value of $8\pi e^{-2/3} = 12.9$.
It is not clear whether these small discrepancies are the result
of the approximations in this paper or uncertainties in Brandt's
numerical work.

\subsection{Ohmic conductor: ac response}

Next, we consider the response of the strip to an ac magnetic
field, $H_{a}(t) = H_{a} \exp(i\omega t)$, so that
$H_{a}(s) = H_{a}/(s-i\omega)$.  When inverting the Laplace
transform, there will be contributions both from the pole at
$i\omega$ and from the square-root branch cut along the negative-$s$
axis.  The branch cut contribution decays as $t^{-1/2}$; since we
are concerned here with the steady-state behavior, we will neglect this
term, keeping only the pole contribution.  Writing
$j_{\rm edge}(\bar{x},t) = j_{\rm edge}(\bar{x},\omega) \exp(i\omega t)$,
we have for the current near the edge
\begin{equation}
j_{\rm edge}(\bar{x},\omega) = {H_{a} \over 2\pi d}
 \left( {\omega \pi a d \over 8 D} \right)^{1/2}
   {\cal F}_{2} \left( {d\omega \bar{x} \over 2 D} \right),
\label{edge4}
\end{equation}
where the ac scaling function ${\cal F}_{2} (u)$ is given by
\begin{equation}
{\cal F}_{2} (u) = {\sqrt{2} \over \pi} \int_{0}^{\infty}
   {e^{-g(y) - i u y  + i\pi/4}  \over y^{1/2} (y^2 +1)^{3/4}} \, dy.
\label{edge5}
\end{equation}
The scaling function is defined so that
${\rm Re}{\cal F}_{2}(0) = {\rm Im}{\cal F}_{2}(0)=1$.
The real and imaginary parts of the ac scaling function are
plotted in Fig. \ref{ac_scaling_fig}.  The real part of the scaling function
has a maximum at $u=0.232$ ($\bar{x}\omega = 0.0738\, a/\tau$ in conventional
units),
with ${\rm Re}{\cal F}_{\rm 2,max} = 1.0787$; the imaginary part changes sign
at
$u=2.43$ ($\bar{x}\omega = 0.773\, a/\tau$ in conventional units).  These
results
are once again very close to Brandt's numerical results
\cite{brandt941,brandt942}.
The uniform approximation to the current density is
\begin{eqnarray}
j_{\rm unif}(x,\omega) & = & {H_{a}\over 2\pi d}
\left\{ - {x\over (1 - x^2)^{1/2}}
  + \left[ {1 \over 2(1-x) } \right]^{1/2}
  - \left[ {1 \over 2(1+x) } \right]^{1/2} \right\} \nonumber \\
   & & \qquad + j_{\rm edge}(1+x,\omega) - j_{\rm edge}(1-x,\omega).
\label{edge6}
\end{eqnarray}

Using the small-$\epsilon$ expansion of the magnetization,
Eq.\ (\ref{moment2}), we can also calculate the high frequency
magnetization; the result is
\begin{equation}
m(\omega) = 1 - {6\over \pi^{3}} {\ln( 8\pi e^{\gamma - 5/3} i \omega \tau)
            \over i \omega \tau} + O(\omega^{-2}),
\label{momega}
\end{equation}
which is quite similar to the numerical result obtained by
Brandt \cite{brandt941}.  The constants differ slightly;
here the prefactor is $0.194$, while Brandt obtained $2/\pi^{2} = 0.203$;
for the constant inside the log, Brandt obtained $16.2$, compared to
the present value of $8\pi e^{\gamma - 5/3} = 8.45$.

\section{Extensions of the method}

With minor modifications it is also possible to treat two related
problems, the current distribution in a thin superconducting disk
in a perpendicular field, and the current distribution in a
thin strip in the presence of an applied current.  Rather than
discussing these cases in detail, only a brief sketch of the results
will be provided.

\subsection{Disk geometry}

Rather than a strip we now have a superconducting disk centered at the origin
of the $x-y$ plane, of radius $a$.  The current density and the vector
potential are both in the $\hat{\phi}$ direction.  Using a Green's
function method similar to that in Sec. II, the $z$-component of the
magnetic field inside the disk satisfies
\begin{equation}
H_{z}(r,z=0,t) = H_{a}(t) + 2 d (P) \int_{0}^{a} P(r,r') j(r',t)\,dr',
\label{disk1}
\end{equation}
where the kernel $P(r,r')$ is given by \cite{brandt942}
\begin{equation}
P(r,r') = {K(k) \over r + r'} + {E(k)\over r'-r},\quad
             k^2 = {4rr'\over (r+r')^2},
\label{disk2}
\end{equation}
where $K(k)$ and $E(k)$ are complete elliptic integrals of the first and
second kind.   We again Laplace transform,  use the constitutive
relation  between $j$ and $A$, and go to the dimensionless
variables $r'=r/a$, $f(r,s) = j(r,s)/(H_{a}(s)/\pi^2 d)$, and
$\epsilon(s) = 2 \lambda_{\rm eff}/a$, to arrive at the following
integro-differential equation:
\begin{equation}
-\epsilon {1\over r} {\partial  \over \partial r}  [rf(r,s)]
    = {\pi\over 2} + {1\over \pi} (P) \int_{0}^{1} P(r,r') f(r',s) dr'.
\label{disk3}
\end{equation}
The outer solution is \cite{CKP,brandt942}
\begin{equation}
f_{0}(r) = {r\over (1-r^2)^{1/2}},
\label{disk4}
\end{equation}
which has a square root singularity at the edge.  Near the edge we
construct the inner solution by defining the inner variables
$R=(1-r)/\epsilon$, $F(R) = \epsilon^{1/2} f(r)$; for small
$\epsilon$ the kernel behaves as
\begin{equation}
P(1-\epsilon R, 1-\epsilon R') = {1\over \epsilon} {1\over R-R'}
                     + O(\ln\epsilon).
\label{disk5}
\end{equation}
After expanding $F(R;\epsilon)$ in $\epsilon$, we find that the
lowest order term satisfies Eq.\ (\ref{inner4}), so that the inner solution is
the same as before.  Finally, the inner and outer solutions are
matched as before; the uniform solution is then
\begin{equation}
f_{\rm unif}(r,s) = {r\over (1-r^2)^{1/2}} - {1\over [2(1-r)]^{1/2}}
     + \epsilon(s)^{-1/2} F_{0}[(1-r)/\epsilon(s)].
\label{disk6}
\end{equation}

The magnetic moment of the disk is
\begin{eqnarray}
M(s)& =& \pi\int_{0}^{a} r^{2} j(r,s)\, dr \nonumber \\
    & = &  {2 a^3 H_{a} \over 3\pi d} m(s),
\label{disk7}
\end{eqnarray}
where the dimensionless moment in the disk geometry is
\begin{equation}
m(s) =  {3 h(s) \over 2} \int_{0}^{1} r^2 f(r,s) \, dr.
\label{disk8}
\end{equation}
Using the uniform approximation, this becomes
\begin{equation}
m(s) = h(s) \left\{ 1 - {4\sqrt{2} \over 5} + {3\over 2\sqrt{2\pi}}
  \int_{0}^{\infty} {e^{-g(\epsilon y)} \left[ y^2 - 2y + 2 - 2 e^{-y}\right]
     \over y^{7/2} ( 1 + \epsilon^2 y^2)^{3/4} } \right\}.
\label{disk9}
\end{equation}
For small $\epsilon$ the integral can be expanded using the same method
as for the strip (see Appendix C), with the result
\begin{equation}
m(s) = h(s) \left[ 1 - {2\sqrt{2} \over \pi} \ln\left({4 e^{\gamma-5/3}
      \over \epsilon}\right) \epsilon + O(\epsilon^2)\right].
\label{disk10}
\end{equation}
For an Ohmic conductor $\epsilon(s) = 2D/ads = 1/\pi\tau s$, with
$\tau = ad/2\pi D$.  After inverting the Laplace transform, we find that
after a jump in the magnetic field, the magnetization for small $t$ is
\begin{equation}
m(t) = 1 - {2\sqrt{2} \over \pi^2} (t/ \tau) \ln\left( 4\pi e^{-2/3}
     t/ \tau\right),
\label{disk11}
\end{equation}
while the ac magnetization at high frequencies is
\begin{equation}
m(\omega) = 1 - {2\sqrt{2} \over \pi^2}
{\ln({4\pi e^{\gamma-5/3} i\omega\tau}) \over i\omega \tau}.
\label{disk12}
\end{equation}
The behavior is essentially the same as for the strip, with slightly
different constants.  Similar behavior has been found by Brandt
\cite{brandt942}
in his numerical studies. He finds a prefactor of $3/\pi^2 = 0.304$, compared
to the present value of $0.286$;  for the constant inside the log,
he obtains $11.3$, compared to our $4.23$.  As in the strip geometry, the
source of the discrepancy is unclear.

\subsection{Current distribution in the presence of an applied current}

The effect of an applied transport current is to modify the $O(1)$
outer solution, which now becomes
\begin{equation}
f_{0}(x) = - {x - f_{a}\over (1-x^2)^{1/2}} ,
\label{applied1}
\end{equation}
where $ f_{a}= I(s)/(H_{a}(s) a /2)$, with $I(s)$ the total current
(not the current density) in the strip.   Writing the outer solution
in terms of the inner variable $X$, we see that the matching conditions
are
\begin{equation}
f_{0}(X) = \pm ( 1 \pm  f_{a}) { 1\over (2\epsilon X)^{1/2}},
\label{applied2}
\end{equation}
with the $+$ corresponding to the left edge and the $-$ to the right edge.
The solution to the inner problem is the same as before.  After matching the
inner and outer solutions, we have for the uniform solution
\begin{eqnarray}
f_{\rm unif} (x,s)& = & - {x - f_{a}\over (1-x^2)^{1/2}}
    + {1 - f_{a} \over [2(1-x)]^{1/2}}
           - {1 + f_{a} \over [2(1+x)]^{1/2}}  \nonumber \\
       & & \qquad + \epsilon(s)^{-1/2} \left\{ (1+ f_{a})
F_{0}[(x+1)/\epsilon(s)]
           -  (1 - f_{a}) F_{0}[(1-x)/\epsilon(s)] \right\}.
\label{applied3}
\end{eqnarray}
With this expansion it is possible to study the time-dependent response,
just as in the zero current case.

\section{Long time behavior after a jump in the applied field}

All of the previous sections have been concerned with the short-time or
high frequency response of a strip to an applied field.  In this section we
will
treat the long-time relaxation of the current density in an Ohmic strip
after a jump in the perpendicular field.   We will again use the method of
matched asymptotic expansions, but in a slightly different form.

We start with the equation of motion for the magnetic field, Eq.\
(\ref{diffeq5}),
and differentiate with respect to time $t$.  The time derivative of the
magnetic
field can be related to the current density for an Ohmic conductor
with conductivity $\sigma$ through
$\partial H_{z}/\partial t = - \partial E_{y}/\partial x = - \sigma^{-1}
\partial j/\partial x$.  Since the applied field is a step function,
$\partial H_{a}(t)/\partial t = H_{a} \delta(t)$.   Therefore, for
$t>0$ the current density satisfies (using $a$ as the unit of length)
\begin{equation}
-{1\over a \sigma} {\partial j(x,t) \over \partial x} = 2 d (P) \int_{-1}^{1}
    {1\over x'-x} {\partial j(x',t) \over \partial t}\, dx'.
\label{longtime1}
\end{equation}
Following Brandt \cite{brandt941,brandt942}, write the current density
as an eigenfunction expansion:
\begin{equation}
j(x,t) = {H_{a} \over 2\pi d} \sum_{n} c_{n} \psi_{n}(x) e^{-t/\tau_{n}},
\label{longtime2}
\end{equation}
where the the relaxation times $\tau_{n}= a d/2D\lambda_{n}$ are related
to the eigenvalues $\lambda_{n}$, which follow from the solution of
\begin{equation}
{d\psi_{n}(x)\over dx} =  {\lambda_{n} \over \pi} (P) \int_{-1}^{1}
{\psi_{n}(x')
              \over x'-x}\, dx'.
\label{longtime3}
\end{equation}
This equation is similar to the homogeneous version of the integral equation
for the current density, Eq.\ (\ref{current2}), but with an important sign
difference on the left hand side.  As a result, we can expect the
eigenfunctions
to be oscillatory, rather than decaying, in the middle of the strip.
The long time behavior of the current density will be controlled by the
smallest eigenvalue $\lambda_{0}$, which produces the longest relaxation
time $\tau_{0}$.

We can develop an asymptotic analysis of the eigenvalue spectrum by first
assuming that $\lambda_{n}\gg 1$, so that $1/\lambda_{n}$ serves as our small
parameter.  The consistency of this assumption should be checked at the
end of the calculation.  As before, we break the problem up into an
outer problem in the middle of the strip, and two inner problems near
each of the two edges.  First we treat the outer problem.
Define the outer variables $X_{o} = \lambda_{n} x$, $\Psi_{n}^{(o)}(X_{o}) =
\psi_{n}(x)$.
Then the integral equation for the outer function is
\begin{equation}
{d \Psi_{n}^{(o)}(X_{o}) \over dX_{o}} =
{1 \over \pi} (P) \int_{-\lambda_{n}}^{\lambda_{n}}
       {\Psi_{n}^{(o)}(X_{o}') \over X_{o}'-X_{o}}\, dX_{o}'.
\label{longtime4}
\end{equation}
Taking $\lambda_{n}\rightarrow \infty$, we then have an integro-differential
equation
which relates the derivative of a function to its Hilbert transform.
The solutions are $\cos(X_{o})$ and $\sin(X_{o})$; however, in the absence
of an applied current the current density must be odd in $x$, so
the physically acceptable solution is
\begin{equation}
\Psi_{n}^{(o)}(X_{o}) = A_{n}  \sin(X_{o}),
\label{longtime5}
\end{equation}
with $A_{n}$ a constant which can in principle depend upon $n$.
This outer solution must be matched onto the inner solution, which we turn to
next.

Let's first consider the inner problem at the left edge.  Define the
inner variables $X_{i} = \lambda_{n} (1+x)$ and $\Psi_{n}^{(i)}(X_{i}) =
\psi_{n}(x)$.
Writing Eq.\ (\ref{longtime3}) in terms of the inner variables, and taking
$\lambda_{n}\rightarrow \infty$, we have
\begin{equation}
{d \Psi_{n}^{(i)}(X_{i}) \over dX_{i}} =
{1 \over \pi} (P) \int_{0}^{\infty}
       {\Psi_{n}^{(i)}(X_{i}') \over X_{i}'-X_{i}}\, dX_{i}'.
\label{longtime6}
\end{equation}
which is once again an integral equation of the Wiener-Hopf type.  To solve,
introduce an unknown function $\tilde{G}(X_{i})$ such that
$\tilde{G}(X_{i})=0$
for $X_{i}>0$, and introduce the complex Fourier transforms
\begin{equation}
\tilde{\Phi}_{+}(k) =\int_{0}^{\infty} \Psi_{n}^{(i)}(X_{i}) e^{i k X_{i}}
dX_{i},
\label{longtime7}
\end{equation}
\begin{equation}
\tilde{G}_{+}(k) = \int_{0}^{\infty} G(X_{i}) e^{i k X_{i}} dX_{i},
\label{longtime8}
\end{equation}
such that $\tilde{\Phi}_{+}(k)$ is analytic for ${\rm Im}(k)>-\beta$ and
$\tilde{G}_{+}(k)$ is analytic for ${\rm Re}(k) <\alpha$, for some
$\alpha>\beta$.  Fourier transforming Eq.\ (\ref{longtime6}), we then obtain
\begin{equation}
\tilde{G}_{+}(k) - \Psi_{n}^{(i)}(0) = i k \tilde{K}(k) \left({k^2 - k_{0}^{2}
\over
                                   k^{2} + \delta^{2}}\right)
\tilde{\Phi}_{+}(k),
\label{longtime9}
\end{equation}
where
\begin{equation}
\tilde{K}(k) =  {k^{2} + \delta^{2} \over k^2 - k_{0}^{2}} \left[ 1 -
   {1\over (k^2 + \delta^2)^{1/2}} \right],
\label{longtime10}
\end{equation}
with $\delta$ a small parameter which is taken to zero at some convenient point
of the calculation, and $k_{0} = (1-\delta^2)^{1/2}$.  The kernel $\tilde{K}$
has been constructed so that it is free of zeros in the strip $-\delta<{\rm
Im}(k)<\delta$,
and $\tilde{K}(k)\rightarrow 1$ as $|k|\rightarrow \infty$.  The kernel
can be factored into the quotient form $\tilde{K}= \tilde{K}_{+}/\tilde{K}_{-}$
using the
general factorization procedure (see Appendix A), with the result that
\begin{equation}
\tilde{K}_{+}(k) = [\tilde{K}(k)]^{1/2} e^{i\tilde{\varphi}(k)},
\label{longtime10a}
\end{equation}
\begin{equation}
\tilde{\varphi}(k) = - {k\over\pi} \int_{0}^{\infty}
{\ln[\tilde{K}(x)/\tilde{K}(k)]
          \over x^2 - k^2 } \, dk.
\label{longtime11}
\end{equation}
With the factorization,  Eq.\ (\ref{longtime9}) can be written as
\begin{equation}
(k-i\delta) \tilde{K}_{-}(k)[ \tilde{G}_{+}(k) - \Psi_{n}^{(i)}(0) ]
  = i k  \left({k^2 - k_{0}^{2} \over k + i\delta}\right) \tilde{K}_{+}(k)
    \tilde{\Phi}_{+}(k).
\label{longtime12}
\end{equation}
Both sides are analytic in their respective regions of analyticity; analytic
continuation allows us to set both sides equal to an entire function $E(k)$.
To choose $E(k)$, we require that $\Psi_{n}^{(i)}(0)$ be finite, so that
$\tilde{\Phi}_{+}(k)\sim 1/k$ for large $k$.  This can only be achieved by
taking $E(k) = B_{n} k/2^{1/2}$, with $B_{n}$ a constant (the $2^{1/2}$ has
been added to
simplify some of the resulting expressions).  Therefore,
\begin{equation}
\tilde{\Phi}{+}(k)  = {B_{n}\over 2^{1/2}} {k + i\delta \over
                      i(k^2 - k_{0}^2) \tilde{K}_{+}(k)}.
\label{longtime13}
\end{equation}
Now we invert the Fourier transform, with the integration path passing above
the poles on the real axis.  Closing the contour in the lower half plane,
we pick up contributions from the poles and from a branch cut which runs
along the ${\rm Im}(k)<0$ axis, with the final result
\begin{equation}
\Psi_{n}^{(i)}(X_{i}) = - B_{n} \cos( X_{i} - \pi/8) + B_{n}\psi_{\rm
cut}(X_{i}),
\label{longtime14}
\end{equation}
where $\psi_{\rm cut}(X_{i})$ is the contribution from the branch cut, which
is $O(X_{i}^{-3/2})$  for large $X_{i}$.

We must now match together our outer solution, Eq.\ (\ref{longtime5}), and our
inner solution, Eq.\ (\ref{longtime14}).   Taking the outer limit of the inner
solution, and rewriting $X_{o}$ and $X_{i}$ in terms of $x$, we find that
\begin{equation}
- B_{n} \cos[ \lambda_{n}(1+x) - \pi/8] = A_{n} \sin(\lambda_{n} x).
\label{longtimematch}
\end{equation}
This is only satisfied if $\lambda_{n} - \pi/8 = (n + 1/2) \pi$ and
$A_{n} = (-1)^{n} B_{n}$.  Therefore, the eigenvalue spectrum is
\begin{equation}
\lambda_{n} = {5\pi \over 8} + n\pi,\qquad n=0,1,2,\ldots.
\label{spectrum}
\end{equation}
The same matching procedure must also be carried out at the right edge, and
the analysis is identical.  From the inner and outer solutions we can construct
a uniform solution, which is
\begin{equation}
\psi_{n,{\rm unif}} (x) = \sin \left[ (n + 5/8)\pi x\right] +
         (-1)^{n} \left\{ \psi_{\rm cut}[\lambda_{n}(1+x)]
            -  \psi_{\rm cut}[\lambda_{n}(1-x)] \right\},
\label{longtimeunif}
\end{equation}
where an overall constant has been dropped.

This asymptotic expansion should be accurate for large $n$.  To compare these
results to Brandt's numerical work \cite{brandt941,brandt942}, first note that
Brandt calculates $\Lambda_{n} = \lambda_{n}/\pi$, so we find $\Lambda_{0} =
5/8=0.625$.
Brandt obtains the numerical value of $\Lambda_{0} = 0.638$, which is
within 2\% of the result of our asymptotic analysis.  The approximation
only improves for large $n$, so our result appears to be quite accurate
for all $n$.  This due in part to the fact that $\lambda_{0} = 5\pi/8$
is large enough for the asymptotic analysis to be effective.
With regard to the eigenfunctions, for large $n$ the cut contributions become
less significant (their contribution is localized near the edges), and so
$\psi_{n}(x) = \sin \left[ (n + 5/8)\pi x\right]$ becomes an accurate
approximation
to the eigenfunctions.  Based on his numerical results, Brandt quotes a similar
result, but with $5/8$ replaced by $1/2$;  our result should provide a better
approximation, and even appears to resemble the numerical result for $n=0$.
The cut contributions have derivatives which diverge logarithmically  at the
edges, similar to the numerical solutions.

\section{Discussion and summary}

By applying the method of matched asymptotic expansions to the
integro-differential
equation for the current density in the strip, Eq. (\ref{current2}), we have
been
able to derive a uniform approximation for the current density, which has been
used to study the nonequilibrium response of the current in the strip, as well
as the ac current density and the magnetization.  Most of the effort has gone
into understanding the response
of an Ohmic strip.  However, the method is easily generalized to more
complicated
dispersive conductivities $\sigma(s)$; the only difficulties arise in inverting
the
Laplace transform to obtain the temporal response.  For the purely Ohmic
response,
we found that after a jump in the perpendicular magnetic field the current
propagates
in from the edges at a constant velocity, in contrast
to the longitudinal case, where the current propagates diffusively
\cite{brandt941}.
The difference is due to the demagnetizing effects in the perpendicular
geometry;
initially the field lines must bend around the edges of the sample, resulting
in a large magnetic ``pressure'' which drives the current into the sample.

A number of simplifying assumptions were made in order to make this problem
analytically tractable.  The most important is the assumption of linear
response.
For many type-II superconductors, however, the current-voltage characteristics
are highly nonlinear due to the collective pinning of the flux lines.
In the perpendicular geometry there has recently been some progress in
incorporating
pinning (nonlinear response) into calculations of the current and field
patterns
for thin superconducting strips \cite{zeldov94,schuster94}.  It is possible
that
the asymptotic methods used in this paper would be useful for studying the
nonequilibrium, nonlinear response in the perpendicular geometry. A second
assumption
is that the current in the strip does not vary in the $y$-direction, so that
we have an essentially one-dimensional problem.  It would be interesting to
include small variations of the current along the $y$-direction in order to
determine
the stability of the current fronts which enter after a jump in the
perpendicular
field; this problem might also be amenable to the type of analysis discussed
in this paper.

\acknowledgments
I would like to thank Dr. Chung-Yu Mou for helpful discussions and for his
assistance
in preparing several of the figures. This work was supported by   NSF Grant DMR
92-23586,
as well as by an  Alfred P. Sloan Foundation Fellowship.

\appendix

\section{Decomposition of the kernel $K$}

We want to decompose the kernel $K(k)$ given in Eq. (\ref{kernel}) into
$K_{+}(k)/K_{-}(k)$, with $K_{+}$ analytic in the upper half plane and
$K_{-}$ analytic in the lower half plane.
Since $K(k)$ is free of zeros in the strip $-\delta < {\rm Im}(k)<\delta$
and approaches 1 as $k\rightarrow \pm \infty$,
$\ln K(k)$ is analytic in this strip
and approaches 0 as $k\rightarrow\pm\infty$.  We can therefore perform
an additive decomposition of
\begin{equation}
\ln K(k) = \ln K_{+}(k)  - \ln K_{-}(k)
\label{decomp3}
\end{equation}
by writing\cite{CKP}
\begin{equation}
\ln K_{+}(k) = {1\over 2\pi i} \int_{-\infty - i\alpha}^{\infty - i\alpha}
          {\ln K (z) \over z - k}\, dz,
\label{decomp4}
\end{equation}
with $\alpha>\delta$; there is an analogous expression for $K_{-}(k)$.
This integral will be analytic for ${\rm Im}(k)>-\delta$.
Now if $k$ can be taken to be real, and the integration path coincides with
the real axis (indented to pass under the pole at $k$), then
\begin{equation}
\ln K_{+}(k) = {1\over 2} \ln K(k)
      + {k\over \pi i} \int_{0}^{\infty} { \ln[K(x)/K(k)]
          \over x^{2} - k^{2} } \, dx,
\label{decomp5}
\end{equation}
where $K(-x)=K(x)$ has been used to simplify the integral.
Therefore, our decomposition of $K=K_{+}/K_{-}$
is (letting $\delta\rightarrow 0$)
\begin{equation}
K_{\pm}(k) =  \left[ 1 + {1\over |k|}\right]^{\pm 1/2} e^{i\varphi(k)},
\label{simple2}
\end{equation}
with
\begin{equation}
\varphi(k) = {{\rm sgn}(k)\over \pi  } \int_{0}^{\infty} \ln\left[
       {u(|k|+1)\over |k| u +1 } \right] { du \over u^{2} -1}.
\label{simple3}
\end{equation}
The last integral may be rewritten in a more convenient form by first
differentiating with respect to $k$, integrating with respect to $u$,
and finally integrating with respect to $k$:
\begin{equation}
 \varphi(k) =  {{\rm sgn}(k)\over \pi } \int_{0}^{|k|}
 {\ln u \over 1- u^{2}}\, du + {\pi\over 4} {\rm sgn}(k).
\label{simple4}
\end{equation}
This integral can be expressed in terms of dilogarithm functions,
but this is not particularly useful for our purposes.

For large $|k|$, $\varphi(k)$ has the asymptotic expansion
\begin{equation}
\varphi(k) \sim  + {1\over \pi} {\ln( e|k|) \over k}
           + O(k^{-3}),
\label{phi1}
\end{equation}
while for small $k$, $\varphi(k)$ has the series expansion
\begin{equation}
\varphi(k) =   {\pi \over 4} {\rm sgn}(k) -{1\over \pi} \ln(e/|k|) k
                + O(k^{3}).
\label{phi2}
\end{equation}

\section{Small $X$ behavior of $F_{0}(X)$}

In this Appendix the behavior of $F_{0}(X)$ for small $X$ will
be derived by using the method of matched asymptotic expansions
to derive a uniformly valid expansion for the integrand of
$F_{0}(X)$.  Call the integrand in Eq. (\ref{rotate1})
$I(y,X)$:
\begin{equation}
I(y,X) = {e^{-Xy - g(y)} \over y^{1/2} (y^2 + 1)^{3/4} } .
\label{asympt1}
\end{equation}
Expanding for small $X$,
\begin{equation}
I_{i}(y,X) = {e^{ - g(y)} \over y^{1/2} (y^2 + 1)^{3/4} }
     - { y^{1/2} e^{ - g(y)} \over (y^2 + 1)^{3/4} } X + O(X^2).
\label{asympt2}
\end{equation}
This constitutes our inner expansion, which breaks down when $Xy = O(1)$.
To derive the outer expansion, define an outer variable $Y = Xy$,
rewrite $I(y,X)$ in terms of $Y$, and expand to lowest order in
$X$:
\begin{eqnarray}
I_{o}(Y,X)  & = & {e^{-Y - g(Y/X)} \over
                   Y^{1/2} (Y^2 + X^2)^{3/4} } X^2 \nonumber \\
            & = & {e^{-Y} \over Y^2} X^{2} + O(X^{3}).
\label{asympt3}
\end{eqnarray}
In order to match the two expansions, express the inner expansion
$I_{i}(y,X)$ in terms of the outer variable $Y$, and expand for
small $X$:
\begin{eqnarray}
I_{i}(Y/X, X)& = & {e^{ - g(Y/X)} \over Y^{1/2}(Y^2 + X^2)^{3/4} } X^{2}
     - { Y^{1/2} e^{ - g(Y/X)} \over  (Y^2 + X^2)^{3/4} } X^{2}
    \nonumber \\
  & = & \left[ {1\over Y^{2} } - {1\over Y} \right] X^{2} .
\label{asympt4}
\end{eqnarray}
On the other hand, if we express the outer expansion $I_{o}(Y,X)$ in
terms of the inner variable $y$, we have
\begin{eqnarray}
I_{o}(Xy, X)& = &{e^{-X y} \over y^2}    \nonumber\\
   & = &  {1\over y^2} - {X \over y} .
\label{asympt5}
\end{eqnarray}
We see that the 1 term outer expansion of the 2 term inner expansion
is equal to the 2 term inner expansion of the 1 term outer expansion,
in agreement with the {\it van Dyke matching principle} \cite{vandyke}.
To obtain the uniform expansion, add the inner and outer expansions,
and subtract the overlap:
\begin{eqnarray}
I_{\rm unif}(y,X)&= & {e^{ - g(y)} \over y^{1/2} (y^2 + 1)^{3/4} }
     - { y^{1/2} e^{ - g(y)} \over (y^2 + 1)^{3/4} } X
    +  {e^{-Xy} \over y^2}
    -{1\over y^2} + {X \over y} + O(X^2)   \nonumber \\
  &= & {e^{ - g(y)} \over y^{1/2} (y^2 + 1)^{3/4} }
     - { y^{1/2}  [e^{ - g(y)}- 1] \over (y^2 + 1)^{3/4} } X \nonumber \\
   & & \qquad  +  {e^{-Xy} -1 \over y^2}
     + {X \over y} - {X y^{1/2}  \over (y^2 + 1)^{3/4}}
     + O(X^2).
\label{asympt6}
\end{eqnarray}
To obtain the small $X$ behavior of $F_{0}(X)$, we can integrate
$I_{\rm unif}(y,X)$ on $y$.  From the arguments given in Sec. IV.C,
we know that $F_{0}(0) = (\pi/2)^{1/2}$, so the first integral is
\begin{equation}
\int_{0}^{\infty} {e^{ - g(y)} \over y^{1/2} (y^2 + 1)^{3/4}}\, dy = \pi.
\label{asympt7}
\end{equation}
The second integral must be evaluated numerically, with the result
\begin{equation}
-X \int_{0}^{\infty} { y^{1/2} [e^{ - g(y)}- 1] \over (y^2 + 1)^{3/4} }\, dy
       = -0.7457\, X.
\label{asympt8}
\end{equation}
The last three integrals require some care.  The third integral is
logarithmically divergent for small $y$, so integrate down to a
cutoff $a$, and take $a\rightarrow 0$ at a later point:
\begin{equation}
\int_{a}^{\infty} {e^{-Xy} -1 \over y^2}\, dy =
    - (1-\gamma) X + X \ln(a) + X\ln(X) + O(a),
\label{asympt9}
\end{equation}
with $\gamma = 0.5772\ldots$.
The last two integrals are logarithmically divergent for large $y$
(the two logs will cancel), so integrate up to $A$ and then take
$A\rightarrow \infty$.  For the fourth integral we then have
\begin{equation}
X \int_{a}^{A} {dy \over y}  = X[\ln(A) - \ln(a)].
\label{asympt10}
\end{equation}
The fifth integral converges for small $y$, so we can extend the
lower limit of integration to 0.  By integrating by parts, we can
extract the leading $\ln (A)$ behavior; there is one remaining
integral which must be calculated numerically, with the result
\begin{equation}
-X \int_{0}^{A} { y^{1/2}  \over (y^2 + 1)^{3/4}}\, dy = - X[ \ln(A)
                    + \ln(2)  - 0.4388].
\label{asympt11}
\end{equation}
Adding Eqs. (\ref{asympt7})--(\ref{asympt11}) together, we see that
the dependence upon $a$ and $A$ drops out, as it should, and we
finally obtain
\begin{eqnarray}
F_{0}(X) & = & {1\over (2\pi)^{1/2}} \int_{0}^{\infty} {e^{-Xy - g(y)} \over
         y^{1/2} (y^2 + 1)^{3/4} }\, dy \nonumber \\
  & = & {1\over (2\pi)^{1/2}}\left[\pi - 1.4228  X  + X\ln X + O(X^2)\right].
\label{asympt12}
\end{eqnarray}
This is slightly different from Eq. (37) of Larkin and
Ovchinnikov \cite{larkin72}, which in our notation is
\begin{eqnarray}
F_{0}^{\rm Larkin}(X) & = &  {1\over (2\pi)^{1/2}}\left[\pi
    - \ln(4 e^{-\gamma})  X  + X\ln X + O(X^2)\right] \nonumber \\
 & = & {1\over (2\pi)^{1/2}}\left[\pi - 0.8091  X
           + X\ln X + O(X^2)\right].
\label{asympt13}
\end{eqnarray}
The difference between these two expressions becomes significant
for values of $X$ near 1. For instance, numerically we find
$F_{0}(1) = 1.732$, while Eq. (\ref{asympt12}) gives $1.718$, a difference
of only $0.8\%$; the Larkin and Ovchinnikov expression gives $2.332$,
an error of $35\%$.  The reason for the discrepancy is not clear,
since the derivation of their result appears not to have been published.

\section{Small $\epsilon$ behavior of the magnetization}

In this Appendix we will determine the small-$\epsilon$ behavior
of the dimensionless magnetization, $m(s)$, given by Eq.\ (\ref{moment1}),
using the method of matched asymptotic expansions.  The derivation is quite
analogous to derivation of the small $X$ behavior of $F_{0}(X)$ which
was discussed in Appendix B. Call the integrand in Eq.\ (\ref{moment1})
$J(y,\epsilon)$:
\begin{equation}
J(y,\epsilon) = { e^{-g(\epsilon y)} [ (y+1)e^{-2y} + y -1] \over
           y^{5/2} [1 + \epsilon^{2} y^2]^{3/4} } .
\label{expand1}
\end{equation}
Expand for small-$\epsilon$, using the small-$x$ behavior of $g(x)$
given in Eq.\ ({g(y)}), to obtain the inner expansion:
\begin{equation}
J_{i}(y,\epsilon) = { (y+1)e^{-2y} + y -1 \over y^{5/2}}
  \left[ 1 +  (y/ \pi) \ln(e/\epsilon y)\epsilon  + O(\epsilon^2)\right].
\label{expand2}
\end{equation}
This expansion breaks down when $\epsilon y = O(1)$.  To derive the outer
expansion, define an outer variable $Y= \epsilon y$, rewrite
$J(y,\epsilon)$ in terms of $Y$, and expand to lowest order in
$\epsilon$:
\begin{equation}
J_{o}(Y,\epsilon) = {e^{-g(Y)} \over Y^{3/2} (1+Y^2)^{3/4} } \epsilon^{3/2}
                    + O(\epsilon^{5/2}).
\label{expand3}
\end{equation}
To match the two expressions, write the inner expansion $J_{i}(y,\epsilon)$
in terms of the outer variable $Y$ and expand for small $\epsilon$:
\begin{equation}
J_{i}(Y/\epsilon,\epsilon)  = Y^{-3/2}\left[ 1 +
              (Y/ \pi) \ln(e/Y) \right] \epsilon^{3/2}.
\label{expand4}
\end{equation}
Next, take the outer expansion $J_{o}(Y,\epsilon)$ and write it in terms of
the inner variable $y$ and expand for small $\epsilon$:
\begin{equation}
J_{o}(\epsilon y,\epsilon) =  y^{-3/2} \left[ 1 +
 (y/ \pi) \ln(e/\epsilon y)\epsilon \right].
\label{expand5}
\end{equation}
Again, we see that the 2 term inner expansion of the 1 term outer expansion
is equal to the 1 term outer expansion of the 2 term inner expansion
\cite{vandyke}.  To obtain the uniform expansion (i.e., an expansion
valid for arbitrary $y$ and small-$\epsilon$), add the
inner and outer expansions, and subtract the overlap:
\begin{eqnarray}
J_{\rm unif}(y,\epsilon) & = & { (y+1)e^{-2y} + y -1 \over y^{5/2}}  +
 (1/ \pi) \ln(e/\epsilon y) { (y+1)e^{-2y}  -1 \over y^{3/2}}
                            \epsilon \nonumber \\
 & & \qquad + {e^{-g(\epsilon y)} -1 \over y^{3/2} (1+\epsilon^2 y^2)^{3/4} }
     +  y^{-3/2} \left[ {1\over (1 + \epsilon^2 y^2)^{3/4}} -1 \right].
\label{expand6}
\end{eqnarray}
We must now integrate $J_{\rm unif}(y,\epsilon)$ over $y$. For the first
term we have
\begin{equation}
\int_{0}^{\infty} { (y+1)e^{-2y} + y -1 \over y^{5/2}}\, dy
          = {2(2\pi)^{1/2}\over 3}.
\label{expand7}
\end{equation}
For the second term we have two integrals:
\begin{equation}
{\ln(e/\epsilon)\over \pi} \int_{0}^{\infty} { (y+1)e^{-2y}  -1
  \over y^{3/2}}\, dy = - {3  \over (2\pi)^{1/2}}\epsilon \ln(e/\epsilon),
\label{expand8}
\end{equation}
\begin{equation}
- {\epsilon\over \pi} \int_{0}^{\infty} \ln y { (y+1)e^{-2y}
   -1 \over y^{3/2}} \, dy
     = -\left({2\over \pi}\right)^{1/2}\left[ - 4 + {9\over 2} \ln 2
     + {3\over 2} \gamma\right]\epsilon .
\label{expand9}
\end{equation}
The fourth and fifth integrals are both $O(\epsilon^{1/2})$, as can be seen
by rescaling the integration variable.  The fourth integral was
performed numerically, with the result
\begin{equation}
\epsilon^{1/2} \int_{0}^{\infty} {e^{-g(x)} -1 \over
  x^{3/2} (1+x^2)^{3/4} }\, dx = 2 \epsilon^{1/2},
\label{expand10}
\end{equation}
where the factor of 2 was determined to an accuracy of 1 part in $10^{8}$.
The last integral is
\begin{equation}
\epsilon^{1/2} \int_{0}^{\infty}  x^{-3/2} \left[ {1\over
   (1 + x^2)^{3/4}} -1 \right]\, dx = - 2\epsilon^{1/2}.
\label{expand11}
\end{equation}
We see that for all purposes the last two integrals sum to zero, although
this has not been proven analytically.  Collecting together the
other terms, we finally have
\begin{equation}
\int_{0}^{\infty}  { e^{-g(\epsilon y)} [ (y+1)e^{-2y} + y -1] \over
           y^{5/2} [1 + \epsilon^{2} y^2]^{3/4} }
   = {2(2\pi)^{1/2} \over 3} \left[ 1 - {9\over 4\pi}  \ln\left(
       {8 e^{\gamma - 5/3} \over \epsilon}\right)\epsilon
      + O(\epsilon^2)\right].
\label{expand12}
\end{equation}
A numerical evaluation of the integral for $\epsilon=1$ gives
$0.49577$; the expansion at $\epsilon=1$ gives $0.48624$,
an error of about 2\%.   We see that the expansion is quite
accurate even for relatively large values of $\epsilon$.

\begin{figure}
\caption{Illustration of the film geometry considered in
this paper.  The film has a width $2a$ in the $x$-direction, and
a thickness $d$.  The applied field $H_{a}$ is in the $z$-direction.}
\label{geometry}
\end{figure}

\begin{figure}
\caption{Schematic diagram of the outer, inner, and matching
regions used in the small-$\epsilon$ asymptotic analysis.}
\label{schematic}
\end{figure}

\begin{figure}
\caption{The inner solution $F_{0}(X,s)$, calculated from
Eq.\ (\protect\ref{rotate3}) (solid line).  Also
shown are the asymptotic expansions for small $X$ (dotted line) and
large $X$ (dashed line).}
\label{inner_fig}
\end{figure}

\begin{figure}
\caption{Uniform approximation $f_{\rm unif}(x,s)$ to the current density for
$\epsilon=0.1$, from Eq.\ (\protect\ref{uniform}) (solid line).  Since the
current density is odd in $x$ only the current density over half the
strip is shown. For comparison the ideal screening case (the outer
solution $f_{0}(x) = - x/(1-x^2)^{1/2}$ ) is also shown (dashed line).
At the left edge we have $f_{\rm unif}(x=-1) = 3.948$. }
\label{uniform_fig}
\end{figure}

\begin{figure}
\caption{Perpendicular component of the magnetic field $H_{z}(x)$ within the
strip
for $\epsilon=0.1$, from Eq.\ (\protect\ref{field}). The field is even in $x$
so
only half the strip is shown.  Note the weak logarithmic singularity at
$x=-1$.}
\label{field_fig}
\end{figure}

\begin{figure}
\caption{Scaling function ${\cal F}_{1}(u)$ for the current density near an
edge
after the magnetic field is suddenly switched on, from
Eq.\ (\protect\ref{newedge}).  There is a maximum at $u=0.386$, with
${\cal F}_{\rm 1,max}=1.1078$.}
\label{scaling_fig}
\end{figure}

\begin{figure}
\caption{Scaling functions for the response to an ac magnetic
field, from Eq.\ (\protect\ref{edge5}); ${\rm Re}{\cal F}_{2}$ is the solid
line and ${\rm Im}{\cal F}_{2}$ is
the dotted line.  The real part has a maximum at $u=0.232$, with
${\rm Re}{\cal F}_{\rm 2,max} = 1.0787$; the imaginary part changes sign at
$u=2.43$.  The integral which defines the scaling function converges
quite slowly, resulting is some numerical inaccuracies which are reflected in
the
small amplitude oscillations in the plots.}
\label{ac_scaling_fig}
\end{figure}

\begin{table}
\caption{Summary of the primary results. The small parameter for the asymptotic
expansion
is $\epsilon = 2 \lambda_{\rm eff}/a$, where $\lambda_{\rm eff}$ is the
effective
penetration depth of the magnetic field and $a$ is either half the width of a
strip
or the radius of a disk. The strip thickness is $d$, $D=1/4\pi\sigma(0)$ is
the diffusion constant for the magnetic field, and $\tau= ad/2\pi D$ is
a relaxation time.  For a strip, $c_{1}=0.194$,
$c_{2} =8.45$; for a disk, $c_{1}=0.286$, $c_{2}= 4.23$.}
\label{table1}
\begin{tabular}{ll}
Outer solution in center (conformal mapping)&  $f_{0}(x)$      \\
Inner solution near the edges (Wiener-Hopf method)&  $F_{0}(X)$      \\
Uniform solution for current density& $f_{\rm unif}(x,s)$                 \\
Uniform solution for magnetic field&  $H_{z}(x,z=0)$                \\
Time-dependent magnetization&
                     $m(t) = 1-c_{1}(t/\tau)\ln(1.526 c_{2} \tau/t)+O(t^2)$\\
Ac magnetization&
$m(\omega)= 1 - c_{1}\ln(c_{2} i\omega \tau)/(i\omega \tau) + O(\omega^{-2})$\\
Scaling of current at edge after jump in the field&
   $ t^{-1/2}{\cal F}_{1}(ad\bar{x}/2Dt) $          \\
Velocity of current propagation after jump in the field&  $v=0.772 D/d$ \\
Scaling of ac current at edge&
  $\omega^{1/2} {\cal F}_{2}(ad\bar{x}\omega/2D)$   \\
Fundamental relaxation time for a strip &  $\tau_{0} = (8/10\pi) ad/D = 0.255
ad/D$   \\
\end{tabular}
\end{table}


\begin{references}

\bibitem[*] {atdemail} Electronic address: atd2h@virginia.edu

\bibitem{brandt941} E. H. Brandt, Phys. Rev. B {\bf 49}, 9024 (1994).

\bibitem{brandt942} E. H. Brandt, Phys. Rev. B {\bf 50}, 4034 (1994).

\bibitem{vandyke} M. van Dyke, {\it Perturbation Methods in Fluid Mechanics}
(Parabolic Press, Stanford CA, 1975).

\bibitem{bender} C. M. Bender and S. A. Orszag, {\it Advanced Mathematical
Methods for Scientists and Engineers} (McGraw-Hill, New York, 1978).

\bibitem{dorsey94} A. T. Dorsey, Ann. Phys. {\bf 233}, 248
(1994); J. C. Osborn and A. T. Dorsey, Phys. Rev. B {\bf 50}, 15 961 (1994).

\bibitem{chapman94} S. J. Chapman, Quart. Appl. Math.
(to be published).

\bibitem{larkin72}  A. I. Larkin and Yu. N. Ovchinnikov, Sov. Phys. JETP
{\bf 34}, 651 (1972).

\bibitem{CKP} G. F. Carrier, M. Krook, and C. E. Pearson,
{\it Functions of a Complex Variable} (Hod Books, Ithaca, New York, 1983),
pp. 376--432.

\bibitem{zeldov94} E. Zeldov, J. R. Clem, M. McElfresh, and M. Darwin,
Phys. Rev. B {\bf 49}, 9802 (1994).

\bibitem{schuster94} T. Schuster, H. Kuhn, E. H. Brandt, M. Indenbom,
M. R. Koblischka, and M. Konczykowski, Phys. Rev. B {\bf 50}, 16 684 (1994).

\end{references}
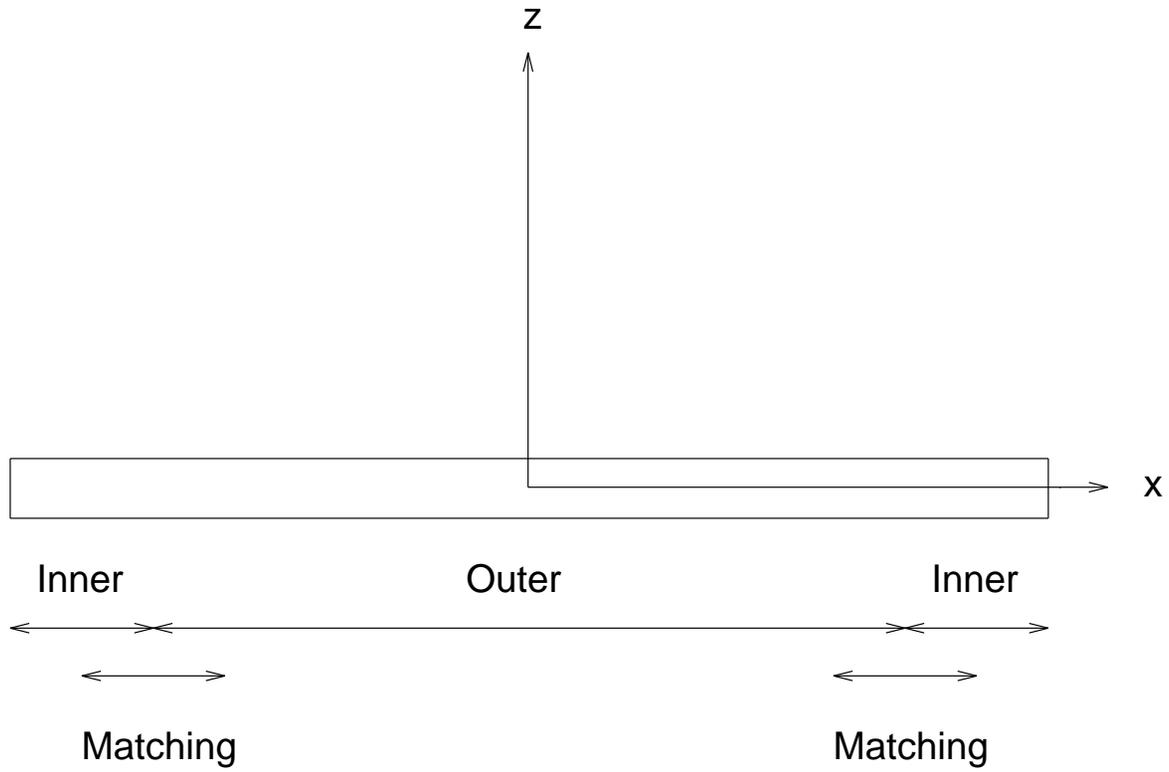
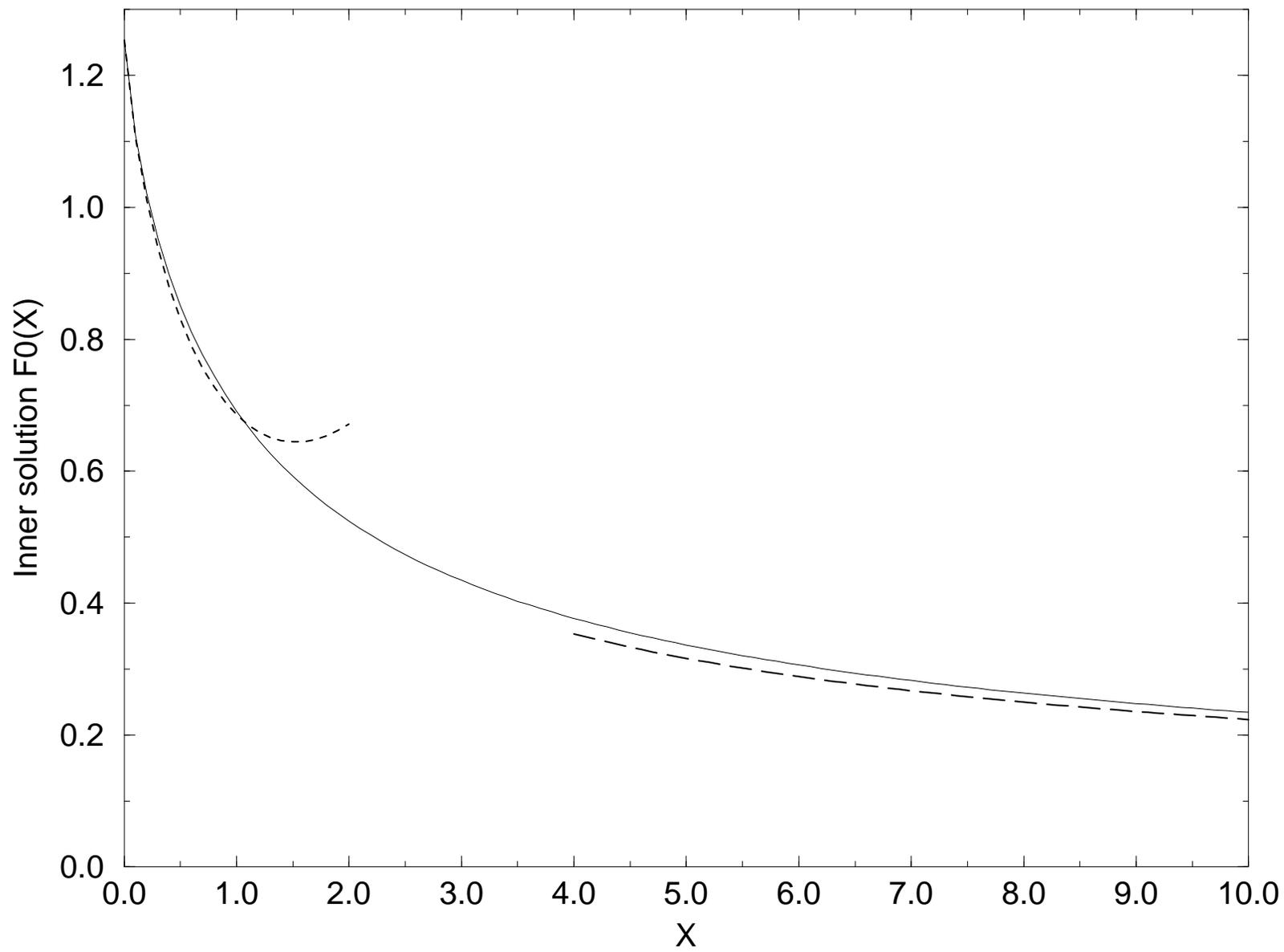
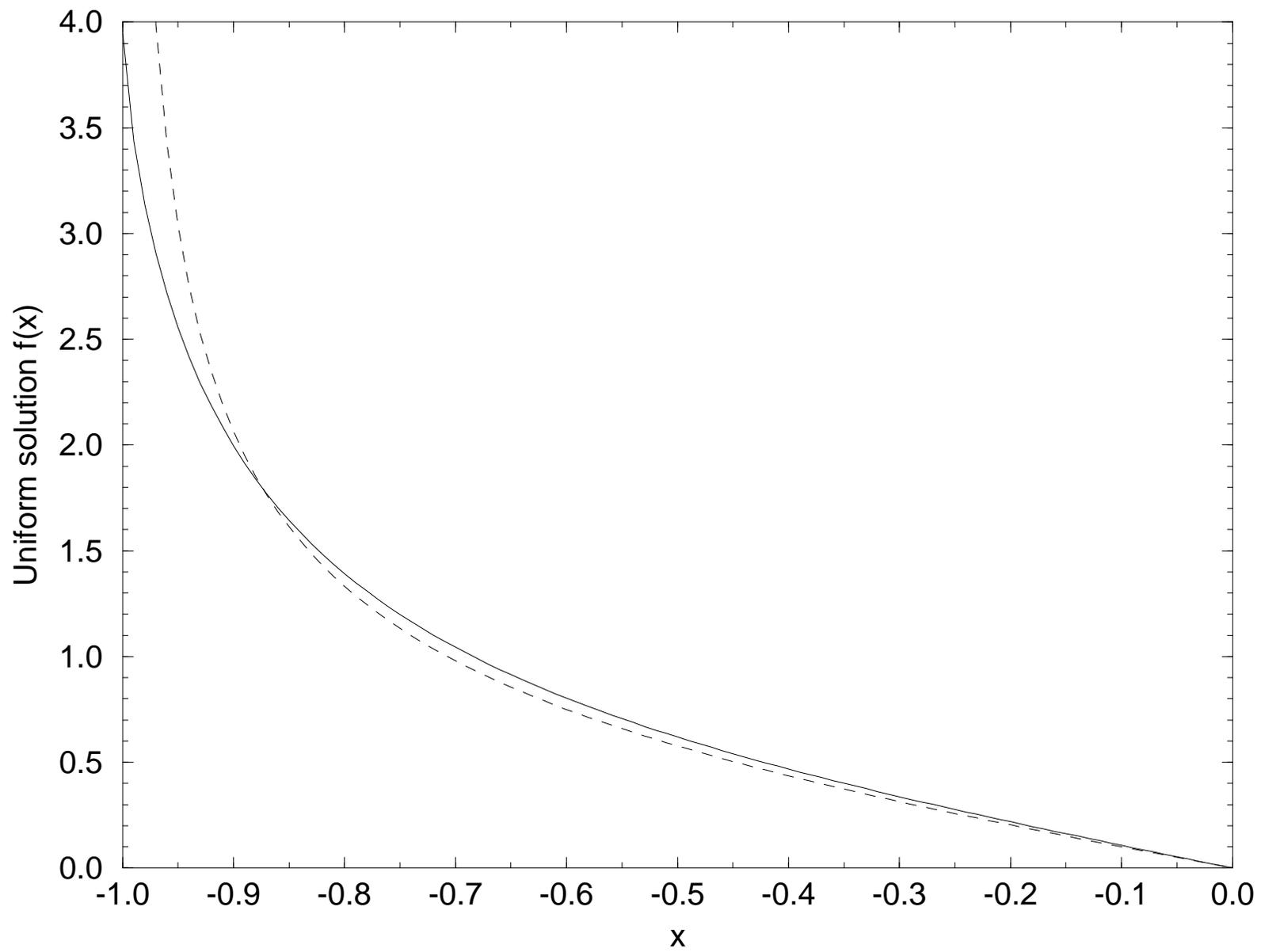
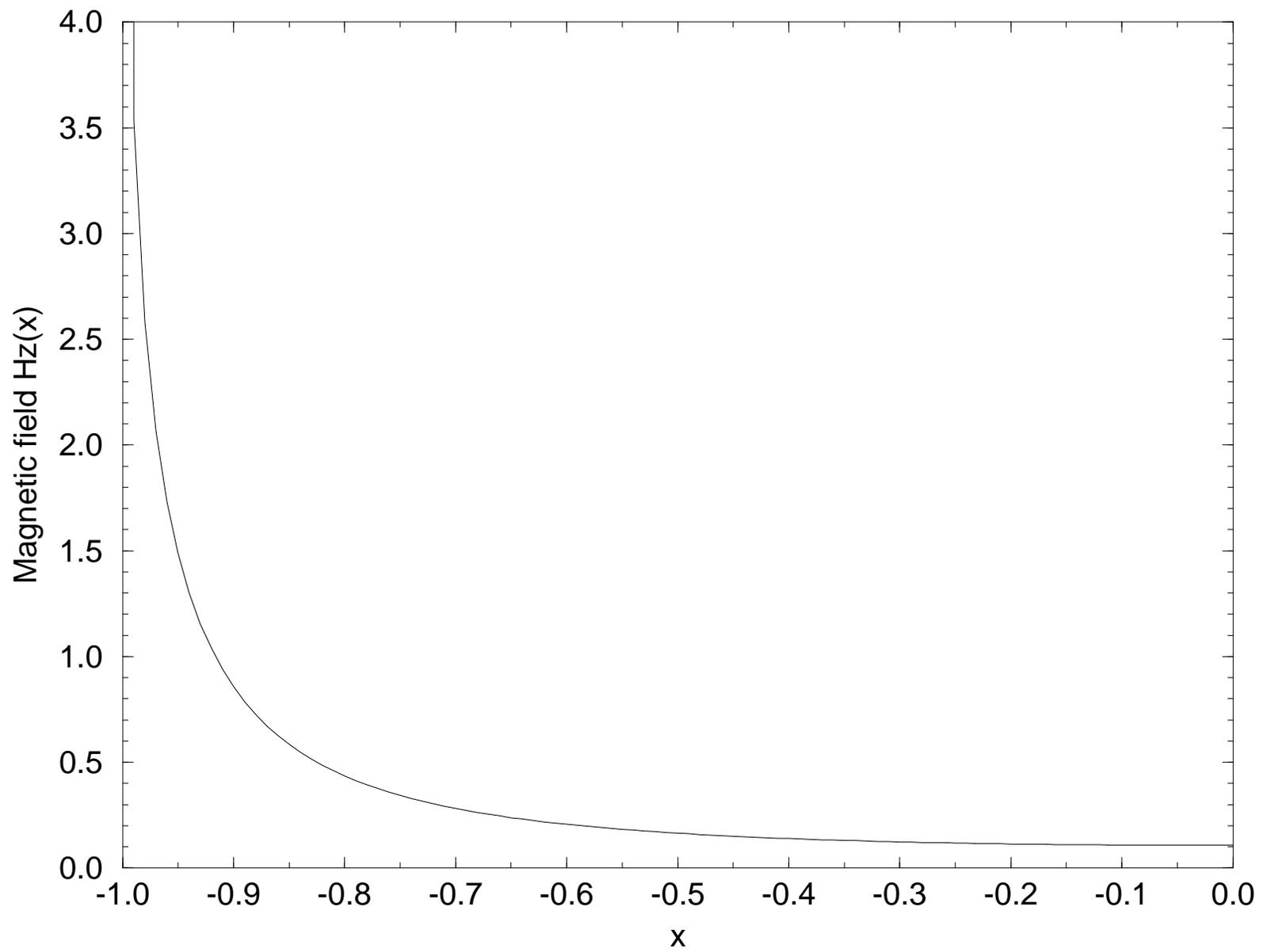
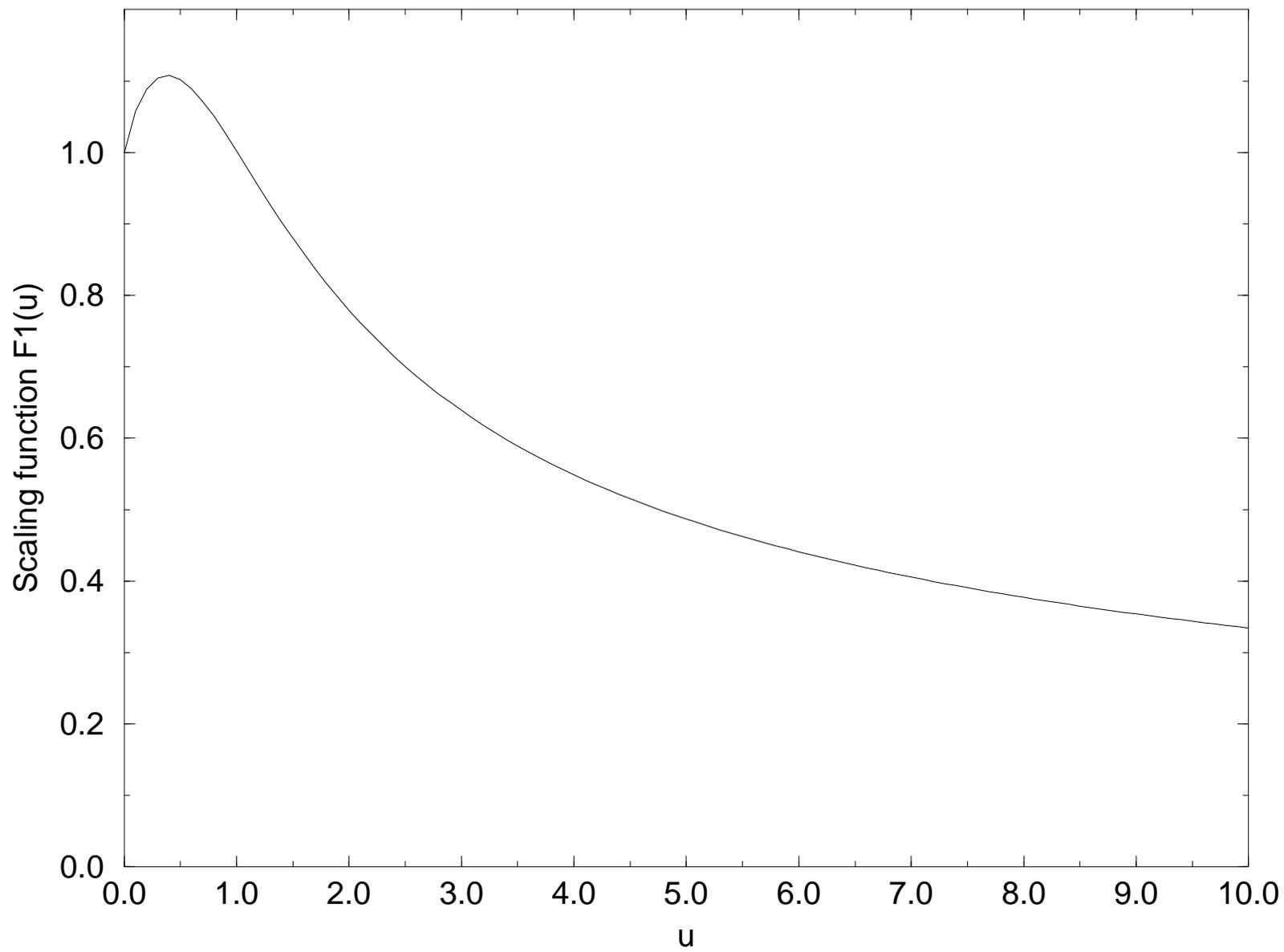
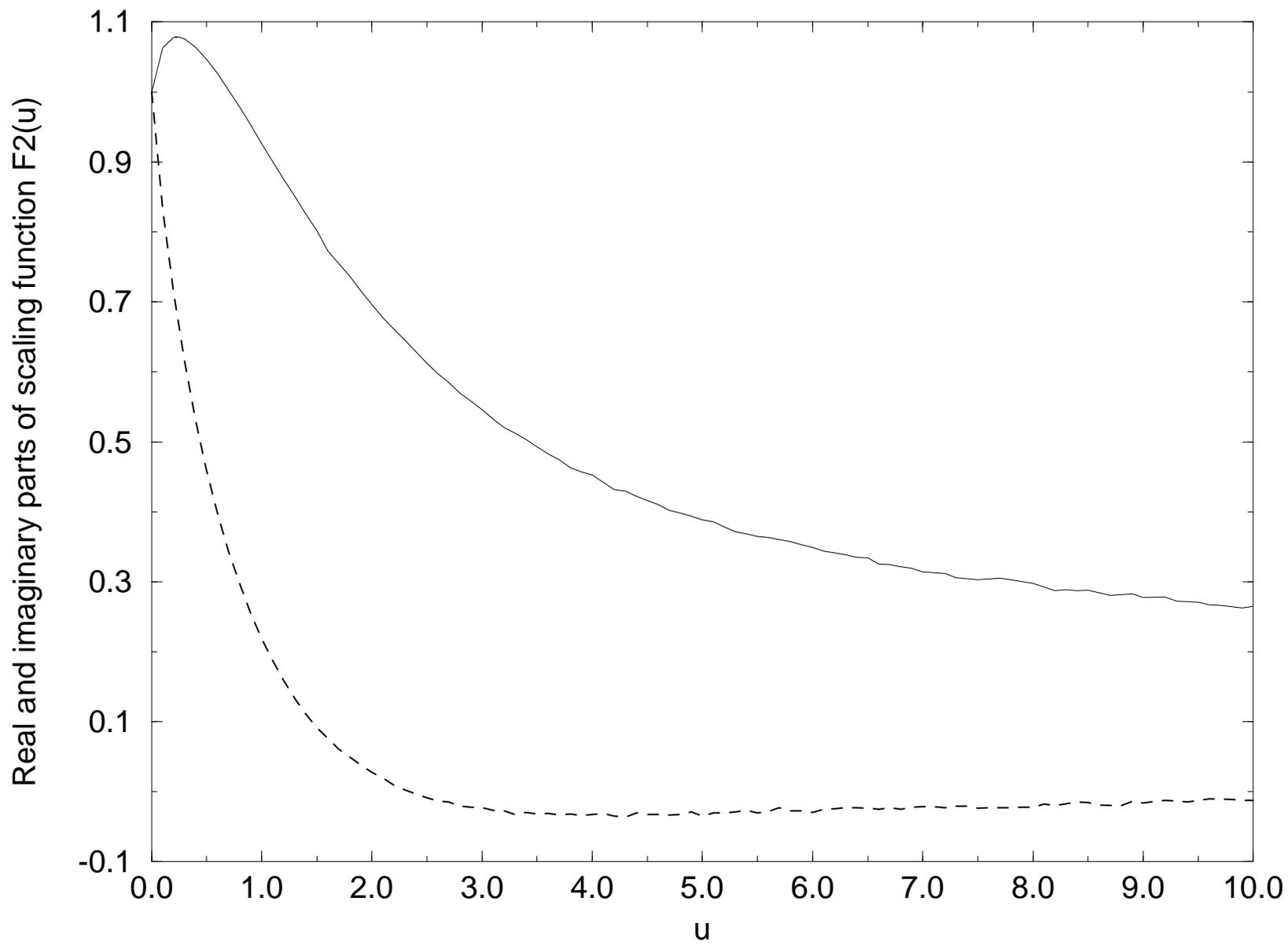
\end{document}